\documentclass[12pt]{article}  

\newcommand{\samplespace}{\mathcal{Z}}

\newcommand{\parameterpriormeasure}{\Pi}
\newcommand{\locationspriormeasure}{\Omega}

\newcommand{\Zm}{Z}  
\newcommand{\numlayers}{H}

\newcommand{\supptitle}{\LARGE Supplementary Material for ``\Paste{title}''}

\usepackage{standalone} 
\usepackage{clipboard}
\usepackage{xr}

\newcommand{\supplement}{%
	\setcounter{section}{0}%
	\renewcommand{\thesection}{S\arabic{section}}%
	\renewcommand{\theHsection}{supp.S\arabic{section}}%
	\setcounter{subsection}{0}%
	\setcounter{table}{0}%
	\renewcommand{\thetable}{S\arabic{table}}%
	\setcounter{figure}{0}%
	\renewcommand{\thefigure}{S\arabic{figure}}%
	\setcounter{equation}{0}%
	\renewcommand{\theequation}{S\arabic{equation}}%
	\setcounter{algorithm}{0}%
	\renewcommand{\thealgorithm}{S\arabic{algorithm}}%
}

\usepackage{etoolbox}
\newbool{arxiv}

\usepackage{soul}

\newcommand{\reff}[1]{\ifbool{arxiv}{\ref{#1}}{\ref*{#1}}}  
\newcommand{\reffsupp}[1]{\reff{#1} of the Supplementary Material} 
\newcommand{\reffmain}[1]{\reff{#1} of the main text} 

\newcommand{\eqreff}[1]{\ifbool{arxiv}{\eqref{#1}}{(\ref*{#1})}}

\usepackage{authblk}

\usepackage{setspace}

\usepackage{scalerel} 
\usepackage[top=1in, bottom=1in, left=1in, right=1in]{geometry}
\usepackage{caption} 
\usepackage{multirow}
\usepackage{graphicx}
\usepackage[authoryear,comma,sectionbib, semicolon]{natbib}
\usepackage{bibunits}
\usepackage{placeins} 
\usepackage{amsmath}	
\usepackage{amsfonts}	
\usepackage{commath}
\usepackage[shortlabels]{enumitem}
\setlist[enumerate,1]{label={(\roman*)}}
\usepackage{amsthm}
\usepackage{tikz}
\usetikzlibrary{positioning}
\usepackage{hyperref}
\usepackage{xcolor}
\definecolor{Red}{rgb}{0.5,0,0}
\definecolor{Blue}{rgb}{0,0,0.5}
\hypersetup{%
	hyperindex = {true},
	colorlinks = {true},
	linktocpage = {true},
	plainpages = {false},
	linkcolor = {Blue},
	citecolor = {Blue},
	urlcolor = {Red},
	pdfstartview = {Fit},
	pdfpagemode = {UseOutlines},
	pdfview = {XYZ null null null}
}

\usepackage{algpseudocode,algorithm,algorithmicx}

\usepackage{tabularx}
\makeatletter
\newcommand{\multiline}[1]{%
	\begin{tabularx}{\dimexpr\linewidth-\ALG@thistlm}[t]{@{}X@{}}
		#1
	\end{tabularx}
}
\makeatother

\newcommand{\proglang}[1]{\texttt{#1}}
\newcommand{\pkg}[1]{{\fontseries{b}\selectfont #1}}

\newcommand{\red}[1]{\textcolor{red}{#1}}

\def\mbf#1{{
		\mathchoice
		{\hbox{\boldmath$\displaystyle{#1}$}}
		{\hbox{\boldmath$\textstyle{#1}$}}
		{\hbox{\boldmath$\scriptstyle{#1}$}}
		{\hbox{\boldmath$\scriptscriptstyle{#1}$}}
}}
\def\vec{\mbf}

\def\d{\textrm{d}} 
\DeclareMathOperator*{\argmax}{arg\,max} 
\DeclareMathOperator*{\argmin}{arg\,min} 

\newcommand{\Gau}{{\text{Gau}}}
\newcommand{\Unif}[2]{{\text{Unif}}(#1, #2)}

\newcommand{\E}{\mathbb{E}} 
\newcommand{\cov}[2]{{\rm cov\!}\left(#1,\, #2\right)} 

\newcommand{\ifootnote}[1]{{\ifthenelse{\isodd{2}}{\footnote{#1}}{}}}
\newcommand{\comment}[1]{{\ifthenelse{\isodd{1}}{\footnote{\red{#1}}}{}}}

\makeatletter
\newenvironment{proof*}[1][\proofname]{\par
	\pushQED{\qed}%
	\normalfont \partopsep=\z@skip \topsep=\z@skip
	\trivlist
	\item[\hskip\labelsep
	\itshape
	#1\@addpunct{.}]\ignorespaces
}{%
	\popQED\endtrivlist\@endpefalse
}
\makeatother

\usetikzlibrary{shapes.multipart}

\tikzset{
  pics/graphstructure/.style args={#1,#2,#3,#4}{
     code={
        \node[main] (1) {$#1$}; 
		\node[main] (2) [above right of=1] {$#2$}; 
		\node[main] (3) [below of=1] {$#3$}; 
		\node[main] (4) [below right of=1, yshift=0.2cm, xshift=0.15cm] {$#4$}; 
            \draw (1) -- (2); 
		\draw (1) -- (3);
		\draw (1) -- (4);
		\draw (3) -- (4);
     }
  }
}

\usetikzlibrary{positioning}

\usetikzlibrary{calc,arrows.meta,positioning}

\usepackage{subcaption}

\usetikzlibrary{decorations.pathreplacing,calligraphy}

\newcommand{\StateSkip}{\vskip 2pt\State}

\newclipboard{quotes}
\setbool{arxiv}{true} 

\title{\LARGE \Copy{title}{Neural Bayes Estimators for Irregular Spatial Data using Graph Neural Networks}} 
\Copy{authors}{%
	\author[1,2]{Matthew Sainsbury-Dale}%
	\author[1]{Andrew Zammit-Mangion}%
	\author[3]{Jordan Richards}%
    \author[2]{Rapha\"el Huser}%
	\affil[1]{School of Mathematics and Applied Statistics, University of Wollongong, Australia}%
	\affil[2]{Statistics Program, Computer, Electrical and Mathematical Sciences and Engineering Division, King Abdullah University of Science and Technology (KAUST), Saudi Arabia}%
	\affil[3]{School of Mathematics, University of Edinburgh, United Kingdom}%
}
\date{}

\begin{document}

\newtheorem{theorem}{Theorem}
\newtheorem{proposition}{Proposition}
\newtheorem{conjecture}{Conjecture}

\begin{singlespace}
\maketitle

\begin{abstract}
\noindent
Neural Bayes estimators are neural networks that approximate Bayes estimators in a fast and likelihood-free manner. Although they are appealing to use with spatial models, where estimation is often a computational bottleneck, neural Bayes estimators in spatial applications have, to date, been restricted to data collected over a regular grid. These estimators are also currently dependent on a prescribed set of spatial locations, which means that the neural network needs to be re-trained for new data sets; this renders them impractical in many applications and impedes their widespread adoption. In this work, we employ graph neural networks (GNNs) to tackle the important problem of parameter point estimation from data collected over arbitrary spatial locations. In addition to extending neural Bayes estimation to irregular spatial data, the use of GNNs leads to substantial computational benefits, since the estimator can be used with any configuration or number of locations and independent replicates, thus amortising the cost of training for a given spatial model. We also facilitate fast uncertainty quantification by training an accompanying neural Bayes estimator that approximates a set of marginal posterior quantiles. We illustrate our methodology on Gaussian and max-stable processes. Finally, we showcase our methodology on a data set of global sea-surface temperature, where we estimate the parameters of a Gaussian process model in 2161 spatial regions, each containing thousands of irregularly-spaced data points, in just a few minutes with a single graphics processing unit. 
\\ 

\end{abstract}
\end{singlespace}

\section{Introduction}\label{sec:introduction}

The computational bottleneck when working with parametric statistical models often lies in making inference on the parameters. A rapidly expanding strand of literature focuses on the use of deep learning and neural networks to facilitate fast likelihood-free inference. Several of these approaches approximate the likelihood function \citep[e.g.,][]{Winkler_2019_likelihood-free_normalising_flows, Papamakarios_2019}, the likelihood-to-evidence ratio \citep[e.g.,][]{Hermans_2020, Thomas_2022_ratio_estimation,  Walchessen_2023_neural_likelihood_surfaces}, the posterior distribution \citep[e.g.,][]{Greenberg_2019, Goncalves_2020, Radev_2022_BayesFlow, Dyer_2022_GNN, Pacchiardi_2022_GANs_scoring_rules}, or both the likelihood function and the posterior \citep[e.g.,][]{Wiqvist_2021, Glockler_2022, Radev_2023_JANA}; see \citet{Zammit_2024_ARSIA} for a recent review. Here, we focus on neural Bayes estimators, which are neural networks that map data to point summaries of the posterior distribution. These estimators are likelihood free, approximately Bayes, and amortised, in the sense that, once trained with simulated data, inference from observed data is (typically) orders of magnitude faster than conventional approaches \citep[for an accessible introduction, see][]{Sainsbury-Dale_2022_neural_Bayes_estimators}. These traits have led to neural Bayes estimators receiving attention in several fields, including population genetics \citep{Flagel_2018}, time-series modelling \citep{Rudi_2020_NN_parameter_estimation}, spatial statistics \citep{Gerber_Nychka_2021_NN_param_estimation, Banesh_2021_neural_estimator_GP, Lenzi_2021_NN_param_estimation, Tsyrulnikov_2024, Sainsbury-Dale_2022_neural_Bayes_estimators, Sainsbury_2025}, and spatio-temporal statistics \citep{Zammit-Mangion_Wikle_2020}. The estimators have also been adapted to settings where data are treated as censored, for example, when fitting certain classes of peaks-over-threshold dependence models for spatial extremes \citep{Richards_2023_censoring}. Despite their promise and growing popularity, neural Bayes estimators for spatial models have, to date, mostly been applied to data collected over a regular grid, as gridded data facilitate the use of parsimonious convolutional neural networks \citep[CNNs;][Ch.~9]{Goodfellow_2016_Deep_learning}. 

The restriction to gridded data is a major limitation in practice. To cater for irregular spatial data, \cite{Gerber_Nychka_2021_NN_param_estimation} propose passing the empirical variogram as input to a multilayer perceptron (MLP). This approach assumes that the empirical variogram is a summary statistic that is highly informative of the parameters. However, while this approach is ideal for Gaussian models, the empirical variogram, which is based on the second moment of the data, is not sufficient for complex non-Gaussian models. More generally, the approach suggested by \cite{Gerber_Nychka_2021_NN_param_estimation} falls into a class of neural approaches that bases estimation on a set of hand-crafted ``good'' (preferably sufficient) summary statistics \citep[see also][]{Creel_2017, Rai_2023}. In practice, finite-dimensional sufficient statistics are not always available and often difficult to construct. Alternatively, one could use an MLP that does not account for the spatial locations of the data; however, ignoring spatial dependence when building a neural Bayes estimator typically leads to poor results \citep{Rudi_2020_NN_parameter_estimation, Sainsbury-Dale_2022_neural_Bayes_estimators}, and such an estimator is again designed for a prescribed set of spatial sample locations, so that the network needs to be re-trained every time the spatial locations change (i.e., for every new data set). Hence, neural Bayes estimation from irregular spatial data remains an open and important problem. 

In this work, we develop amortised neural Bayes estimators for irregular spatial data. Our novel approach involves representing the data as a graph with edges weighted by spatial distance, and then employing graph neural networks \citep[GNNs;][]{Zhang_2019_GCN_review, Zhou_2020_GNN_review, Wu_2021_GNN_review}. GNNs generalise the convolution operation in conventional CNNs to graphical data, and they have recently been used for regression problems in spatial statistics by, for example, \cite{Tonks_2022_GNNs_geostats}, \cite{Zhan_Datta_2023_GNNs_geostats}, and \cite{Cisneros_2023_GNNs_Australian_wildfire}. We also propose a GNN architecture tailored for learning summary statistics that can be expected to be highly informative of spatial dependence parameters; the empirical variogram is in the class of statistics that can be learnt by our GNN. Compared to MLPs, GNNs provide a more parsimonious representation for constructing neural Bayes estimators for irregular spatial data, since the ``graphs'' in GNNs can be used to encode spatial dependence. The explicit modelling of spatial dependence facilitates the learning of a useful mapping between the sample space and the parameter space, and allows the estimator to generalise to unseen spatial configurations \citep{Bronstein_2017, Battaglia_2018}. In particular, a single GNN-based neural Bayes estimator can be used with data collected over any number or configuration of spatial locations, and this means that the often-expensive training stage needs to be performed only once for a given spatial model. In addition to proposing the use of GNNs for the estimation of spatial-model parameters, we also consider several important practical issues: in particular, we show how to construct synthetic spatial data sets for training such an estimator; how to design a suitable architecture to make inference from data from a single spatial field or from multiple replicates of a spatial process; and how to perform rigorous uncertainty quantification in an amortised manner, by training a neural Bayes estimator that approximates marginal posterior quantiles in a way that respects their ordering. Finally, to facilitate the use of GNN-based neural Bayes estimators by practitioners, we incorporate our methodology in the user-friendly software package \pkg{NeuralEstimators} \citep{NeuralEstimators}, which is available in the \proglang{Julia} and \proglang{R} programming languages.
  
The remainder of this paper is organised as follows. In Section~\ref{sec:methodology}, we describe neural Bayes estimation for irregular spatial data using GNNs. In Section~\ref{sec:simulationstudies}, we illustrate the strengths of the proposed approach by way of extensive simulation studies based on Gaussian and max-stable processes. In Section~\ref{sec:application}, we apply our methodology to the analysis of a massive global sea-surface temperature data set. In Section~\ref{sec:conclusion}, we conclude and outline avenues for future research. Supplementary material is also available that contains additional details and figures. Code that reproduces all results in the manuscript is available from \url{https://github.com/msainsburydale/NeuralEstimatorsGNN}.

\section{Methodology}\label{sec:methodology}

In Section~\ref{sec:neuralPointEstimators}, we briefly review neural Bayes estimators. In Section~\ref{sec:irregulardata}, we describe how GNNs may be used to perform neural Bayes inference from irregular spatial data. 
 
\subsection{Neural Bayes estimators}\label{sec:neuralPointEstimators}

The goal of parameter point estimation is to estimate unknown model parameters $\vec{\theta} \in \Theta$ from data $\vec{Z} \in \samplespace$ using an estimator, $\hat{\vec{\theta}} : \samplespace\to\Theta$, where $\samplespace$ is the sample space and $\Theta$ is the parameter space. For notational convenience, we focus on the case where $\samplespace \subseteq \mathbb{R}^n$ and $\Theta \subseteq \mathbb{R}^p$; what we propose also applies to discrete-data and discrete-parameter settings \citep[see, e.g.,][]{Chan_2018}. 
 Point estimators can be constructed intuitively within a decision-theoretic framework. Consider a non-negative loss function, $L(\vec{\theta}, \hat{\vec{\theta}})$, which assesses an estimate $\hat{\vec{\theta}}$ for a given $\vec{\theta}$. An estimator's Bayes risk is its unconditional risk, 
\begin{equation}\label{eqn:risk}
 \int_\Theta \int_{\samplespace}  L(\vec{\theta}, \hat{\vec{\theta}}(\vec{Z}))f(\vec{Z} \mid \vec{\theta}) \d \vec{Z} \d \parameterpriormeasure(\vec{\theta}),  
 \end{equation} 
where $\parameterpriormeasure(\cdot)$ is a prior measure for $\vec{\theta}$ and $f(\cdot \mid \vec{\theta})$ is the probability density function of the data $\vec{Z}$ given $\vec{\theta}$. A minimiser of \eqref{eqn:risk} is 
a \textit{Bayes estimator} with respect to $L(\cdot,\cdot)$ and $\parameterpriormeasure(\cdot)$. 

 Bayes estimators are theoretically attractive, being consistent and asymptotically efficient under mild conditions \citep[Thm.~5.2.4; Thm.~6.8.3]{Lehmann_Casella_1998_Point_Estimation}. Unfortunately, Bayes estimators are typically unavailable in closed form. Recently, motivated by universal function approximation theorems \citep[e.g.,][]{Hornik_1989_FNN_universal_approximation_theorem, Zhou_2018_universal_approximation_CNNs}, neural networks have been used to approximate Bayes estimators. Let $\hat{\vec{\theta}}(\vec{Z}; \vec{\gamma})$ denote a neural network that returns a parameter point estimate from data $\vec{Z}$, with $\vec{\gamma}$ comprising the neural-network parameters. Then, a Bayes estimator may be approximated by $\hat{\vec{\theta}}(\cdot; \vec{\gamma}^*)$, where 
\begin{equation}\label{eqn:optimisation_task}
\vec{\gamma}^*
\equiv 
\underset{\vec{\gamma}}{\mathrm{arg\,min}} \; \frac{1}{K} \sum_{k=1}^K L(\vec{\theta}^{(k)}, \hat{\vec{\theta}}(\vec{Z}^{(k)}; \vec{\gamma})). 
\end{equation} 
 The objective function in \eqref{eqn:optimisation_task} is a Monte Carlo approximation of \eqref{eqn:risk} made using a set $\{\vec{\theta}^{(k)} : k = 1, \dots, K\}$ of parameter vectors sampled from the prior measure $\parameterpriormeasure(\cdot)$ and, for each $k$, data $\vec{Z}^{(k)}$ simulated from $f(\vec{z} \mid  \vec{\theta}^{(k)})$. The optimisation task \eqref{eqn:optimisation_task} is a form of empirical risk minimisation (\citeauthor{Goodfellow_2016_Deep_learning}, \citeyear{Goodfellow_2016_Deep_learning}, pg.~268--269; see also \citeauthor{Xu_Raginsky_2022}, \citeyear{Xu_Raginsky_2022}), and it can be solved efficiently using back-propagation and stochastic gradient descent; moreover, it does not involve evaluation, or even knowledge, of the likelihood function. Note that the use of simulated data in \eqref{eqn:optimisation_task} allows for the construction of arbitrarily large training data sets and, therefore, the use of large, expressive neural networks that are prone to overfitting when trained with small data sets. The fitted neural network given by \eqref{eqn:optimisation_task} approximately minimises the Bayes risk, and is thus called a \textit{neural Bayes estimator} \citep{Sainsbury-Dale_2022_neural_Bayes_estimators}. The procedure is summarised in Algorithm~\ref{alg:NBE}. 

Neural Bayes estimators have a number of strengths. First, once a moderate-to-large computational cost has been paid to complete the optimisation task \eqref{eqn:optimisation_task}, the trained estimator can be applied repeatedly to real data sets at almost no computational cost. These estimators are therefore ideal for settings in which the same statistical model must be fit repeatedly (e.g., online estimation problems), 
in which case the initial training cost is said to be ``amortised'' over time. Due to their amortised nature, they are also amenable to rapid bootstrap-based uncertainty quantification, which is usually considered to be relatively accurate but computationally prohibitive; amortised uncertainty quantification can also proceed via a separate neural Bayes estimator trained to approximate the marginal posterior quantiles (see Section~\ref{sec:UQ}). Finally, for models with an analytically or computationally intractable likelihood function, neural Bayes estimators often provide a significant improvement over popular approximate methods (see Section~\ref{sec:Schlather}), which often justifies their training cost even in single-use cases. 

\begin{algorithm}[!t]
  \caption{Amortised inference using neural Bayes estimators}  \label{alg:NBE}
  \begin{algorithmic}[1]
\smallskip
    \Statex \hspace{-1.75em}\textbf{Training stage} (slow)
\Require{Number of training samples $K$, prior $\parameterpriormeasure(\vec{\theta})$, model $f(\vec{Z} \mid \vec{\theta})$ for the data $\vec{Z}$ given parameters $\vec{\theta}$, neural-network architecture for $\hat{\vec{\theta}}(\cdot; \vec{\gamma})$, loss function $L(\cdot, \cdot)$.} 
    \State  Sample parameters $\vec{\theta}^{(k)} \sim \Pi(\vec{\theta})$ for $k = 1, \dots, K$.
     \StateSkip  Simulate data $\vec{Z}^{(k)} \sim f(\vec{Z} \mid \vec{\theta}^{(k)})$ for $k = 1, \dots, K$.
     \StateSkip  Solve the optimisation task $\vec{\gamma}^* \equiv \mathrm{argmin}_\vec{\gamma} \;
\frac{1}{K} \sum_{k=1}^K L(\vec{\theta}^{(k)}, \hat{\vec{\theta}}(\vec{Z}^{(k)}; \vec{\gamma}))$.
	\StateSkip Return $\hat{\vec{\theta}}(\cdot; \vec{\gamma}^*)$. 
	\vskip 10pt
    \Statex \hspace{-1.75em}\textbf{Assessment stage} (fast)
    \setcounter{ALG@line}{0}
    \State Assess $\hat{\vec{\theta}}(\cdot; \vec{\gamma}^*)$ using simulation-based methods, for instance by analysing the empirical sampling distribution of the estimator and its properties (e.g., bias, variance, etc.). 
    \State If the estimator passes assessment, proceed to the inference stage: otherwise, return to the training stage with a larger value of $K$ and/or a modified neural-network architecture. 
    \vskip 10pt
    \Statex \hspace{-1.75em}\textbf{Inference stage} (fast, repeatable for arbitrarily many observed data sets) 
    \setcounter{ALG@line}{0}
\Require{Observed data $\vec{Z}$.} 
    \State Compute point estimates $\hat{\vec{\theta}} = \hat{\vec{\theta}}(\vec{Z}; \vec{\gamma}^*)$.
    \State Uncertainty quantification for $\hat{\vec{\theta}}$ via bootstrap sampling or by using a second neural Bayes estimator to approximate a set of marginal posterior quantiles (Section~\ref{sec:UQ}). 
  \end{algorithmic}
\end{algorithm}

Specification of a prior distribution is a requirement of Bayesian methods, and most of the usual considerations also apply to neural Bayes estimators. However, there are some important considerations that are specific to neural Bayes estimators (and amortised simulation-based methods more broadly). First, if the estimator is to be re-used for several applications, then one should employ a sufficiently vague prior, since the prior cannot be updated post-training. Second, if the estimator is intended for a single application, an informative prior with compact and relatively narrow support can be useful for reducing the volume of the parameter space that must be sampled from when performing the optimisation task \eqref{eqn:optimisation_task}; in this case, a good approximation of the Bayes estimator can typically be obtained with a smaller value of $K$ in \eqref{eqn:optimisation_task} than that required under a vague prior. 

There is a need for more theory in this emerging field to determine the conditions needed for a trained neural Bayes estimator to be within an `$\epsilon$-ball' of the true Bayes estimator with high probability: conditions on the size of the network in terms of its width and depth, and the number of training samples (i.e., $K$ in \eqref{eqn:optimisation_task}), would be particularly helpful in guiding practitioners. While such theoretical developments are important, they are beyond the scope of this paper. Nevertheless, it is straightforward to empirically assess the performance of a trained neural Bayes estimator; by applying the estimator to many simulated data sets, one can quickly and accurately assess the properties of its sampling distribution. 

As discussed in Section~\ref{sec:introduction}, neural Bayes estimators for spatial data have, to date, mostly been limited to data collected over a regular grid, in which case CNNs may be employed. MLPs can in principle be used for irregular data, but they do not explicitly account for spatial dependence, and they are conditional on a specific configuration of spatial locations. Constructing an estimator based on user-defined summary statistics of fixed dimension is an appealing option, but it is only feasible when easy-to-compute ``near-sufficient'' summary statistics are available. We therefore seek an architecture that parsimoniously models spatial dependence; that is able to yield a statistically efficient estimator by automatically constructing statistics from the full data set that are highly informative of the model parameters; and that can be applied to data collected over arbitrary spatial locations. The next subsection describes how this can be achieved.

\subsection{Neural Bayes estimators for irregular spatial data}\label{sec:irregulardata}

In Section~\ref{sec:GNN:GeneralFramework}, we describe how GNNs may be used to construct neural Bayes estimators for irregular spatial data, where inference is made from a single field. In Section~\ref{sec:trainingdata}, we describe how to account for varying spatial locations. In Section~\ref{sec:replicateddata}, we describe how GNNs may be used to estimate parameters from independent replicates of a spatial process. In Section~\ref{sec:UQ}, we discuss the important task of uncertainty quantification. 

\subsubsection{Inference from a single spatial field}\label{sec:GNN:GeneralFramework}

GNNs are a class of neural networks designed for graphical data, and they have been the subject of reviews by \citet{Zhang_2019_GCN_review}, \citet{Zhou_2020_GNN_review}, and \citet{Wu_2021_GNN_review}. 
GNNs generalise the convolution operation in conventional CNNs and, therefore, they are able to  efficiently extract information on the dependence structure in graphical data. GNNs can also generalise to different graphical inputs (of potentially different sizes, connections, edge weights, etc.), and they can scale well with the graph size, particularly when the graphs are sparse. These properties make GNNs natural candidates for constructing neural Bayes estimators for irregular spatial data, where the spatial data are viewed as a (sparse) graph with edges weighted by a decaying function of spatial distance. In what follows, we assume that we have data $\vec{Z} \equiv (Z_1, \dots, Z_n)'$ observed at locations $S \equiv \{\vec{s}_1, \dots, \vec{s}_n\} \subset \mathcal{D}$, where $\mathcal{D}$ is the spatial domain of interest. 

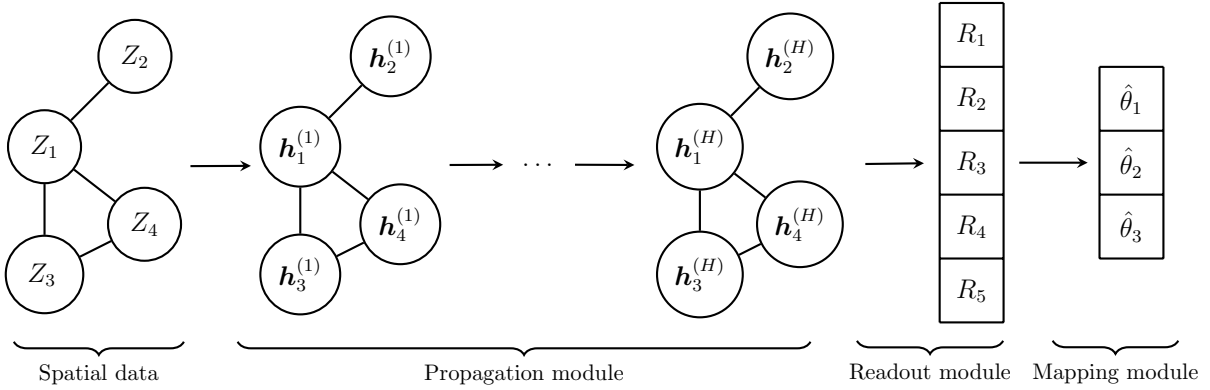
\begin{figure}[t!]
\centering
\begin{tikzpicture}[
roundnode/.style={circle, draw=black!100, minimum size=7mm},
squarednode/.style={rectangle, draw=red!60, fill=red!5, very thick, minimum size=5mm},
scale = 0.85, 
every node/.style={scale=0.85}, 
node distance={20mm}, 
thick, 
main/.style = {draw, circle}, 
every text node part/.style={align=left}
]

\begin{scope}[xshift = -3.5cm]
\draw[local bounding box=Z] (0,0) pic{graphstructure={\;Z_{1}\;, \;Z_{2}\;, \;Z_{3\;}\;, \;Z_{4}\;}};
\end{scope}
\begin{scope}[xshift = 0.5cm]
\draw[local bounding box=h1] (0,0) pic{graphstructure={\vec{h}^{(1)}_{1}, \vec{h}^{(1)}_{2}, \vec{h}^{(1)}_{3}, \vec{h}^{(1)}_{4}}};
\end{scope}
\node at (4.25, -0.28) (dots) {$\dots$};
\begin{scope}[xshift = 6.75cm]
\draw[local bounding box=hL] (0,0) pic{graphstructure={\vec{h}^{(\numlayers)}_{1}, \vec{h}^{(\numlayers)}_{2}, \vec{h}^{(\numlayers)}_{3}, \vec{h}^{(\numlayers)}_{4}}};
\end{scope}

\begin{scope}[xshift = 10.5cm,yshift = -0.75cm]
\draw[step=1cm] (0,-2) grid (1, 3);
\node (V1) at (0.5, 2.5) {$R_1$};
\node (V2) at (0.5, 1.5) {$R_2$};
\node (V3) at (0.5, 0.5) {$R_3$};
\node (V4) at (0.5, -0.5) {$R_4$};
\node (V5) at (0.5, -1.5) {$R_5$};
\end{scope}

\begin{scope}[xshift = 13cm, yshift = -0.25cm]
\draw[step=1cm, yshift=0.5cm] (0,-2) grid (1,1);
\node (theta1) at (0.5, 1) {$\hat{\theta}_1$};
\node (theta) at (0.5, 0) {$\hat{\theta}_2$};
\node (theta3) at (0.5, -1) {$\hat{\theta}_3$};
\end{scope}

\draw[shorten >=0.125cm, shorten <=0.125cm, -stealth, thick] (Z) -- node[above,font=\bfseries] {} (h1);
\draw[shorten >=0.125cm, shorten <=0.125cm, -stealth, thick] (h1) -- node[above,font=\bfseries] {} (dots);
\draw[shorten >=0.3cm, shorten <=0.1cm, -stealth, thick] (dots) -- node[above,font=\bfseries] {} (hL);
\draw[shorten >=0.3cm, shorten <=0.3cm, -stealth, thick] (hL) -- node[above,font=\bfseries] {} (V3);
\draw[shorten >=0.3cm, shorten <=0.3cm, -stealth, thick] (V3) -- node[above,font=\bfseries] {} (theta);

\draw[decorate,decoration={brace,amplitude=5pt,mirror}] 
([shift={(-7.8,-2.6)}]dots.south west) -- ([shift={(-6.0,-2.6)}]dots.south east) 
node[midway,below=5pt] {\footnotesize Spatial data};

\draw[decorate,decoration={brace,amplitude=5pt,mirror}] 
([shift={(-4.3,-2.6)}]dots.south west) -- ([shift={(3.8,-2.6)}]dots.south east) 
node[midway,below=5pt] {\footnotesize Propagation module};

\draw[decorate,decoration={brace,amplitude=5pt,mirror}] 
([shift={(-1.4,-0.46)}]V5.south west) -- ([shift={(0.1,-0.46)}]V5.south east) 
node[midway,below=5pt] {\footnotesize Readout module};

\draw[decorate,decoration={brace,amplitude=5pt,mirror}] 
([shift={(-0.9,-1.4)}]theta3.south west) -- ([shift={(0.4,-1.4)}]theta3.south east) 
node[midway,below=5pt] {\footnotesize Mapping module};

\end{tikzpicture}
\caption{
The architecture of our proposed GNN-based neural Bayes estimator for a single spatial field. The data $\vec{Z} = (Z_1, \dots, Z_n)'$ and their spatial locations $S = \{\vec{s}_1, \dots, \vec{s}_n\}$ are sequentially convolved by the $\numlayers$-layered propagation module into a graph with hidden-feature vectors $\{\vec{h}^{(\numlayers)}_1, \dots,  \vec{h}^{(\numlayers)}_n\}$. Pairs of nodes are determined to be neighbours or not based on spatial proximity and a maximum neighbour count. The readout module summarises this graph into a vector of summary statistics, $\vec{R}$, that is fixed in length irrespective of the size of the input graph. Finally, the mapping module transforms $\vec{R}$ into parameter estimates, $\hat{\vec{\theta}} = (\hat{\theta}_1, \dots, \hat{\theta}_p)'$, where the nonlinear mapping is done using an MLP. 
}\label{fig:GNN-estimator}
\end{figure}

In the context of deep learning, parameter estimation from irregular spatial data constitutes a ``graph-level regression task'', where the entire graph (spatial data) is associated with some fixed-dimensional vector (model parameters) that we wish to estimate. The architecture of a typical GNN used for graph-level regression consists of three modules that are applied sequentially: the propagation module, the readout module, and the mapping module. See Figure~\ref{fig:GNN-estimator} for an illustration of these modules. 

In the \textit{propagation module}, a graph-convolution operator is applied to each node to form a series of hidden feature graphs, which have the same size and structure as the input graph \citep[unless the graph-coarsening technique known as ``local pooling'' is applied between propagation layers; see, e.g.,][]{Mesquita_2020_pooling, Grattarola_2022_pooling}. A large class of propagation modules can be couched in the so-called ``message-passing'' framework \citep{Gilmer_2017_message_passing_GNNs}, where spatial-based convolutions are performed locally on each node (i.e., vertex of the graph) and its neighbours. Information is passed between non-neighbouring nodes by applying local convolutions in successive layers. This approach scales well with the graph size, since only a subset of nodes are considered for each computation, and it allows a GNN to generalise to different graph structures, since the convolutional parameters are shared across the graph. Following \citet{Danel_2020_spatial_GNN}, \citet{Zhang_2021_spatial_GNN}, and \citet{Klemmer_2023_spatial_GNN}, among others, we explicitly incorporate spatial information in our propagation module which, as we show in Figure~\reffsupp{fig:spatial_vs_nonspatial}, is an important design choice. One has flexibility in designing the propagation module; we define ours as
\begin{align}
\vec{h}^{(l)}_{j} &=
 g\Big(
 \vec{\Gamma}_{\!1}^{(l)} \vec{h}^{(l-1)}_{j}
 +
 \vec{\Gamma}_{\!2}^{(l)} \bar{\vec{h}}^{(l)}_{j}
 +
 \vec{b}^{(l)}
 \Big),\label{eqn:propagation1}\\
 \bar{\vec{h}}^{(l)}_{j} &= \sum_{j' \in \mathcal{N}(j)}\tilde{\vec{w}}_j^{(l)}(\vec{s}_j, \vec{s}_{j'}) \odot \vec{\rho}^{(l)}(\vec{h}^{(l-1)}_{j}, \vec{h}^{(l-1)}_{j'}),\label{eqn:propagation2}
\end{align}
 where, for $j=1, \dots, n$ and layers $l = 1, \dots, \numlayers$, $\vec{h}^{(l)}_{j}$ is the hidden-feature vector at location $\vec{s}_j$, $h_{j}^{(0)} \equiv Z_j$, $g(\cdot)$ is a nonlinear activation function applied elementwise, $\vec{\Gamma}_{\!1}^{(l)}$ and $\vec{\Gamma}_{\!2}^{(l)}$ are trainable parameter matrices, $\vec{b}^{(l)}$ is a trainable bias vector, $\mathcal{N}(j)$ denotes the indices of neighbours of $\vec{s}_j$, $\odot$ and $\oslash$ respectively denote elementwise multiplication and division, $\vec{\rho}^{(l)}(\cdot, \cdot)$ is a learnable function detailed below, and 
 \begin{align}
\tilde{\vec{w}}_j^{(l)}(\vec{s}_j, \vec{s}_{j'}) &= \vec{w}^{(l)}(\vec{s}_j, \vec{s}_{j'}) \oslash \sum_{j'' \in \mathcal{N}(j)}\vec{w}^{(l)}(\vec{s}_j, \vec{s}_{j''}),\label{eqn:propagation3}
\end{align}
is a normalised version of a (learnable) spatial weight function $\vec{w}^{(l)}(\cdot, \cdot)$, whose elements are strictly positive. We consider isotropic processes only, and we therefore model $\vec{w}^{(l)}(\vec{s}_j, \vec{s}_{j'}) \equiv \vec{w}^{(l)}(\|\vec{s}_j - \vec{s}_{j'}\|)$ as a function of spatial distance using a combination of spatial basis functions \citep{Cressie_2021_review_spatial-basis-function_models} and an MLP (see Section~\ref{sec:SimulationIntro} for details). 
 We set $\vec{\rho}^{(l)}(\vec{h}^{(l-1)}_{j}, \vec{h}^{(l-1)}_{j'}) = |a^{(l)}\vec{h}^{(l-1)}_{j} - (1 - a^{(l)})  \vec{h}^{(l-1)}_{j'}|^{b^{(l)}}$ for learnable parameters $a^{(l)} \in [0, 1]$ and $b^{(l)} > 0$, and where the absolute-value operation and exponentiation are done elementwise.  
 In Section~\reffsupp{sec:variogram}, we motivate this representation through the lens of the empirical (semi)variogram, which was used as a summary statistic in the context of neural Bayes estimation by \cite{Gerber_Nychka_2021_NN_param_estimation}. In Section~\reffsupp{sec:neighbourhoods}, we investigate several definitions of the neighbourhood in \eqref{eqn:propagation2}. We find that deterministically selecting a subset of $k$ neighbours within a disc of fixed radius $r$ leads to good statistical and computational performance, and that the performance of the estimator is relatively robust to the hyperparameters $k$ and $r$. We therefore adopt this definition throughout. We give further details on our specific choice of architecture in Section~\ref{sec:SimulationIntro}.  
  
In the \textit{readout module}, the graph output from the propagation module is aggregated into a vector of summary statistics, $\vec{R}$, which is fixed in length irrespective of the size and structure of the input graph. We express this readout module as 
\begin{equation}\label{eqn:readout}
\vec{R} = \vec{r}(\{\vec{h}^{(\numlayers)}_j : j = 1, \dots, n\}),
\end{equation}
where the readout function, $\vec{r}(\cdot)$, is a permutation-invariant set function, and recall that $n$ denotes the number of spatial locations. Each element of $\vec{r}(\cdot)$ is typically chosen to be a simple aggregation function (e.g., elementwise addition, average, or maximum), but more flexible readout modules have also been proposed in the context of general graph-level regression \citep[e.g.,][]{Zhang_2018_SortPool, Navarin_2019_universal_readout}. In this paper, we use the elementwise average. 
 Note that when modelling nonstationary processes, it may be necessary to define $\vec{r}(\cdot)$ as a combination of a simple aggregation function like the elementwise mean (to obtain an average of locally-computed summary statistics) and a pooling operation that preserves locality \citep[e.g., spatial pyramid pooling;][]{He_2014_spatial_pyramid_pooling}. For many statistical models used in practice, the number of summary statistics required to reach ``near-sufficiency'' for $\vec{\theta}$ is unknown and, in these cases, the dimension of $\vec{R}$ should be chosen to be reasonably large \citep[see][for a discussion]{Zammit_2024_ARSIA}. Since $\vec{R}$ has fixed dimension, a single GNN-based neural Bayes estimator can be used to make inference from data $\vec{Z}$ collected over any number and configuration of spatial locations. 
 
Finally, the \textit{mapping module} maps the summary statistics $\vec{R}$ into parameter estimates, 
\begin{equation}\label{eqn:theta_single_field}
\hat{\vec{\theta}} = \vec{\phi}(\vec{R}; \vec{\gamma}_\phi),
\end{equation}
where $\vec{\phi}(\cdot; \vec{\gamma}_\phi)$ is an MLP parameterised by $\vec{\gamma}_\phi$. Note that the final activation function in $\vec{\phi}(\cdot; \vec{\gamma}_\phi)$ determines the range of each element of $\hat{\vec{\theta}}$ (e.g., identity activations allow for unconstrained estimates, exponential or softplus activations ensure positive estimates, etc.). Our estimator can thus be viewed as a nonlinear mapping of summary statistics $\vec{R}$, which are themselves nonlinear mappings of the data $\vec{Z}$ and spatial locations $S$. Finally, one may make the estimator a function of both $\vec{R}$ and hand-crafted summary statistics for $\vec{Z}$ \citep[e.g., the empirical variogram;][]{Gerber_Nychka_2021_NN_param_estimation} and $S$ (e.g., Ripley's $K$-function), or local variants of these summary statistics for nonstationary processes. However, since $\vec{R}$ can be expected to approximate well any summary statistic as a continuous function of $\vec{Z}$ and $S$, the choice to include hand-crafted summary statistics is mainly a practical one intended to simplify the learning task. 


\subsubsection{Training the estimator to account for varying spatial locations}\label{sec:trainingdata}

A GNN-based neural Bayes estimator is a function of the spatial locations $S$ at which the data are collected, and it can be applied to data collected over any number and configuration of spatial locations. If one wishes to make inference from a single spatial data set, and this data set is collected before the estimator is constructed, then training data in Algorithm~\ref{alg:NBE} can be simulated using the observed spatial locations, which can be treated as fixed and known. However, to construct an estimator that is approximately Bayes for a large range of spatial configurations (and number $n$ of spatial locations), one requires an estimator that is adaptive to the observed spatial locations. To this end, we propose treating $S$ as a random point pattern (i.e., drawn from a point process). Then, assuming that $S$ is independent of $\vec{\theta}$, the Bayes risk \eqref{eqn:risk} becomes  
\begin{equation}\label{eq:riskv2}
\int_{\mathcal{S}} \int_\Theta\int_{\samplespace}  L(\vec{\theta}, \hat{\vec{\theta}}(\vec{Z}, S))f(\vec{Z} \mid \vec{\theta}, S) \d \vec{Z} \d \parameterpriormeasure(\vec{\theta}) \d \locationspriormeasure(S),
\end{equation}
where $\mathcal{S}$ is the space of all possible spatial configurations and $\locationspriormeasure(\cdot)$ is a distribution for $S$. The number of spatial locations, $n$, is a random quantity whose distribution is implicitly defined by $\locationspriormeasure(\cdot)$. When $S$ is treated as random, an additional step is required in the training stage of Algorithm~\ref{alg:NBE}: spatial locations, $S^{(k)} \sim \locationspriormeasure(S)$, $k = 1, \dots, K$, are sampled before simulating data, $\vec{Z}^{(k)} \sim f(\vec{Z} \mid \vec{\theta}^{(k)}, S^{(k)})$.

The task of choosing a distribution $\locationspriormeasure(\cdot)$ for the spatial locations is simplified by two properties of the proposed framework. First, since we consider isotropic spatial processes, our estimator depends only on spatial distances, and inference can be made using the data with spatial coordinates scaled by a common factor such that they are contained within the unit square; estimates of any range parameters are then scaled back accordingly. A similar strategy can be employed with stationary anisotropic processes, where the estimator depends only on spatial lags. Therefore, for such processes, one need only consider distributions $\locationspriormeasure(\cdot)$ of spatial configurations defined on the unit square. Second, Theorem~\ref{thm:GNNproof} in Appendix~\ref{app:Proof:priormeasureS} shows that if $S$ is independent of $\vec{\theta}$, as is the case in most applications, the Bayes estimator is invariant to the specific choice of $\locationspriormeasure(\cdot)$ among all strictly positive distributions on $\mathcal{S}$. Therefore, besides ensuring positivity, the choice of $\locationspriormeasure(\cdot)$ is theoretically immaterial. In practice, however, the empirical risk function in \eqref{eqn:optimisation_task} is subject to Monte Carlo error that could be large in regions of $\mathcal{S}$ where $\locationspriormeasure(\cdot)$ assigns low probability; the choice of $\locationspriormeasure(\cdot)$ therefore has practical implications. 

\begin{figure*}[t!]
    \centering
    \includegraphics[width = \textwidth]{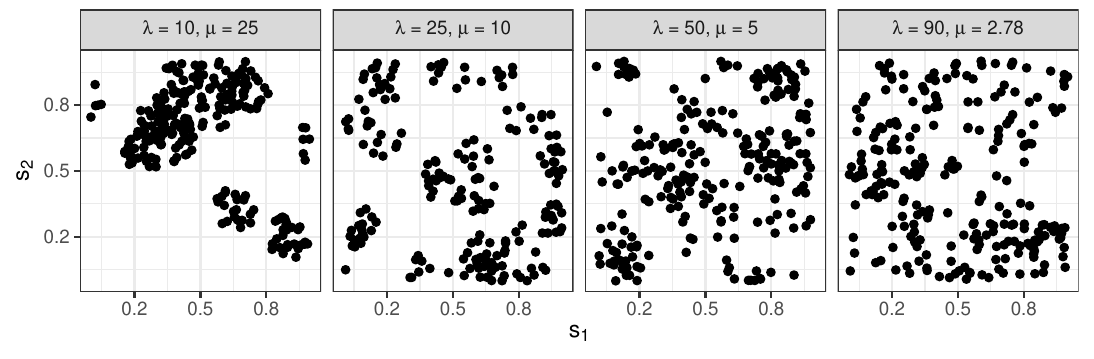}  
    \caption{Realisations from a Matérn cluster process with parent Poisson point process intensity $\lambda$, mean number of daughter points $\mu$, and cluster-disk radius $\delta = 0.1$. Various parameter combinations (see panel titles) are chosen such that the expected number of sampled points, $\lambda\mu$, is fixed to 250. Spatial point processes are useful when constructing training data in order to cover a wide range of spatial configurations.
    }\label{fig:spatialpatterns}
\end{figure*}

If no prior knowledge on the spatial configuration is available, then $\locationspriormeasure(\cdot)$ could be chosen to be reasonably uninformative to produce an estimator that is broadly applicable. Spatial point-process models \citep{Moller_2004_point_patterns, Illian_2008_point_patterns, Diggle_2013_point_patterns} are ideal for this purpose. A convenient point-process model, among many candidates, is the Matérn cluster process \citep[][Ch.~12]{Baddeley_2015_point_patterns}. Simulation from a Matérn cluster process proceeds by first drawing a random point pattern from a homogeneous Poisson point process with intensity $\lambda > 0$; then, with each point in this underlying (unobserved) point pattern serving as the centre of a disk with constant radius $\delta > 0$, a Poisson($\mu$)-distributed number of points are simulated uniformly over each disk. Figure~\ref{fig:spatialpatterns} shows realisations from a Matérn cluster process under several parameter choices. In practice, the parameters $\lambda$, $\mu$, and $\delta$ may be selected by visualising realisations from the cluster process and modifying the parameters until the sampled spatial configurations cover a sufficiently wide range of scenarios (e.g., from sparse to dense, and from highly clustered to approximately uniform). In Section~\ref{sec:simulationstudies}, we show that training under a broad range of spatial configurations allows a GNN-based neural Bayes estimator to perform well irrespective of the locations of the data. 
 
Bayes estimators are generally also a function of the number of spatial locations, $n$, and this must be accounted for if the estimator is to generalise over a wide range of possible sample sizes. To illustrate the role of $\locationspriormeasure(\cdot)$ when $n$ is variable, in Section~\reffsupp{sec:experiment:variablesamplesizes} we consider the case where the sampling process of the spatial locations is known but the specific sample size, $n$, is unknown. We see that treating $n$ as a random variable results in an estimator that performs near-optimally over the values of $n$ for which it was trained; in contrast, an estimator trained with fixed $n$ does not necessarily extrapolate well to observed data sets with different sample sizes, particularly when $n$ is small. In Section~\reff{sec:experiment:variablesamplesizes}, we also consider the case in which inference is required from a specific set of locations, $S_0$. We find that an estimator trained with $S$ treated as random can be just as efficient in terms of the number of simulations required to make accurate inferences with $S_0$, as an estimator trained with $S = S_0$ fixed.

\subsubsection{Inference from independent replicates}\label{sec:replicateddata}

Inference is often made from multiple independent replicates of a spatial field, particularly when modelling spatial extremes, or when working with highly-parameterised or weakly-identifiable models. In this case, we have multiple graphs (of potentially different structures) associated with a single output, which is not a standard problem in the GNN literature. We address this challenge by couching GNNs within the DeepSets framework \citep{Zaheer_2017_Deep_Sets}, which was also employed in the context of neural Bayes estimation for gridded spatial data by \cite{Sainsbury-Dale_2022_neural_Bayes_estimators}. Suppose that we have data from $m$ mutually independent replicates of a spatial process that we collect in $\Zm \equiv \{\vec{Z}_1, \dots,\vec{Z}_m\}$, where the locations, $S_i \equiv \{\vec{s}_{i1}, \dots, \vec{s}_{in_i}\}$, and number of observations, $n_i$, are allowed to vary between realisations $i = 1, \dots, m$. 
Then, DeepSets-based parameter estimates may be evaluated from the data and their spatial locations through
\begin{equation}\label{eqn:DeepSets} 
  \hat{\vec{\theta}} = \vec{\phi}(\vec{T}; \vec{\gamma}_\phi), \quad 
\vec{T}  = \vec{a}\big(\{\vec{R}_{\!i} : i = 1, \dots, m\}\big), 
\end{equation}
where $\vec{R}_{i}$ is the summary statistic for $\vec{Z}_i$ computed using the propagation and readout modules \eqref{eqn:propagation1}--\eqref{eqn:readout},  
 \mbox{$\vec{a}(\cdot)$} is a permutation-invariant set aggregation function (here chosen to be the elementwise average), 
 and \mbox{$\vec{\phi}(\cdot; \vec{\gamma}_\phi)$} is an MLP parameterised by $\vec{\gamma}_\phi$. Figure~\reffsupp{fig:GNN-estimator:DeepSets} illustrates the architecture \eqref{eqn:DeepSets}. Note that, when applied to a single replicate, \eqref{eqn:DeepSets} reduces to the architecture proposed in Section~\ref{sec:GNN:GeneralFramework}. 

The representation \eqref{eqn:DeepSets} has several motivations. First, Bayes estimators are invariant to permutations of independent replicates; estimators constructed from \eqref{eqn:DeepSets} are guaranteed to be permutation invariant. Second, the DeepSets representation is known to have universality properties for continuously differentiable permutation-invariant functions \citep[e.g.,][]{Wagstaff_2021_universal_approximation_set_functions, Han_2019_universal_approximation_of_symmetric_functions}; an estimator constructed in the form of \eqref{eqn:DeepSets} can therefore be expected to approximate well any Bayes estimator that is a continuously differentiable function of the data. Third, \eqref{eqn:DeepSets} may be applied to data sets with an arbitrary number of replicates, $m$, which allows the training cost to be amortised with respect to the number of replicates. Fourth, the Bayes estimator depends on the sample size $m$ and is not, in general, equal to the average of single-replicate Bayes estimates \citep[see][Fig.~2]{Sainsbury-Dale_2022_neural_Bayes_estimators}; the construction \eqref{eqn:DeepSets} allows a neural estimator to approximate the true ``multiple-replicate Bayes estimator''. See \citet{Sainsbury-Dale_2022_neural_Bayes_estimators} for further details on the use of the DeepSets architecture in the context of neural Bayes estimation, and for a discussion on the architecture's connection to conventional estimators. 

\subsubsection{Uncertainty quantification}\label{sec:UQ}

Uncertainty quantification with neural Bayes estimators often proceeds through the bootstrap distribution \citep[e.g.,][]{Lenzi_2021_NN_param_estimation, Richards_2023_censoring, Sainsbury-Dale_2022_neural_Bayes_estimators}. Bootstrap-based approaches are particularly attractive when nonparametric bootstrap is possible (e.g., when the data are independent replicates), or when simulation from the fitted model is fast, in which case parametric bootstrap is also computationally efficient. However, these conditions are not always met in spatial statistics. For example, when making inference from a single spatial field, nonparametric bootstrap is not possible without breaking the spatial dependence structure, and the cost of simulation from the fitted model is often non-negligible (e.g., exact simulation from a Gaussian process model requires the factorisation of an $n\times n$ matrix, where $n$ is the number of spatial locations, which is a task that is $\mathcal{O}(n^3)$ in computational complexity). Further, although bootstrap-based methods for uncertainty quantification are often considered to be fairly accurate and favourable to methods based on asymptotic normality, there are situations where bootstrap procedures are not reliable \citep[see, e.g.,][pg.~6]{Canty_2006}. 

Alternatively, by leveraging ideas from (Bayesian) quantile regression \citep[e.g.,][]{Koenker_1978_quantile_regression, Koenker_2001_quantile_regression, Yu_Moyeed_2001}, one may construct a neural Bayes estimator that approximates a set of marginal posterior quantiles \citep{Fisher_2023_neural_quantile_regression}, which can then be used to construct univariate credible intervals for each parameter. Inference then remains fully amortised since, once the estimators are trained, both point estimates and credible intervals can be obtained with virtually zero computational cost. Posterior quantiles can be targeted by employing the quantile loss function which, for a single parameter $\theta$, is 
\begin{equation}\label{eqn:quantileloss_univariate}
L_\tau(\theta, \hat{\theta}) = (\hat{\theta} - \theta)(\mathbb{I}(\hat{\theta} > \theta) - \tau), \quad \tau \in (0, 1),
\end{equation}
where $\mathbb{I}(\cdot)$ denotes the indicator function. In particular, the Bayes estimator under \eqref{eqn:quantileloss_univariate} is the posterior $\tau$-quantile. When there are $p > 1$ parameters, $\vec{\theta} = (\theta_1, \dots, \theta_p)'$, the Bayes estimator under the joint loss ${L(\vec{\theta}, \hat{\vec{\theta}}) = \sum_{k=1}^p L_\tau(\theta_k, \hat{\theta}_k)}$ is the vector of marginal posterior quantiles since, in general, a Bayes estimator under a sum of univariate loss functions is given by the vector of marginal Bayes estimators (see Theorem~\ref{theorem:additive_loss_functions} in Appendix~\ref{app:Proof:separablelosses}). 
 
 The above approach to uncertainty quantification recasts classical quantile regression from a task of estimating quantiles of a response variable conditional on covariates, to a task of estimating marginal quantiles of $\vec{\theta}$ conditional on data $\vec{Z}$. The use of neural networks in quantile regression dates back to at least \cite{Taylor_2000_quantile_regression}, and more recent applications include, for example, \citet{Cannon_2011_quantile_regression_neural_networks}, \citet{ Xu_2017_quantile_regression_neural_network}, \citet{Pfreundschuh_2018}, \citet{Pasche_Engelke_2022}, \citet{Richards_Huser_2022_PINN_quantile_regression}, and \citet{Zhong_Wang_2023_partially_interpretable_neural_quantile_regression}. A consideration in quantile regression is monotonicity of the estimated quantile functions: the $\tau_1$-quantile should not exceed the $\tau_2$-quantile for any $\tau_2 > \tau_1$. When this fundamental property does not hold, the estimated quantiles curves are said to cross \citep{Bassett_Koener_1982, He_1997_quantile_curves_without_crossing}. The longstanding quantile-crossing problem can be addressed by simply reordering the quantiles after their estimation \citep{Chernozhukov_2010, Alcantara_2023}, or by choosing a functional form for the regression function that ensures monotonicity with respect to $\tau$. In this work, we take the latter approach, explicitly imposing monotonicity through our neural-network design. 

A monotonic neural network \citep[e.g.,][]{Sill_1997_Monotonic_networks, Gupta_2016_monotonic_networks, Cannon_2018} could be used for the quantile network $\vec{q}(\vec{Z}, \tau)$ that takes as input the data $\vec{Z}$ and the desired probability level $\tau$. However, architectures that ensure (partial) monotonicity typically impose constraints on the neural-network parameters and activation functions, which can limit the expressiveness of the neural network and complicate the training procedure \citep{Wehenkel_Louppe_2019_monotonic_networks}. Further, if the relationship between $\tau$ and the $\tau$-quantile is highly nonlinear, a network that takes $\tau$ as input would need to be more complex than one that treats $\tau$ as fixed. Therefore, following \cite{Madrid-Padilla_2022}, we restrict our attention to making inference for a fixed set of probability levels $\{\tau_1, \dots, \tau_T\}$, and employ a separate neural network for each probability level: 
 \begin{align}\label{eqn:quantileneuralnetwork}
\begin{aligned}
\vec{q}^{(\tau_1)}(\vec{Z}) &= \vec{v}^{(\tau_1)}(\vec{Z}),\\
\vec{q}^{(\tau_t)}(\vec{Z}) &= \vec{v}^{(\tau_1)}(\vec{Z}) + \sum_{j=2}^t g(\vec{v}^{(\tau_j)}(\vec{Z})), \quad t = 2, \dots, T, 
\end{aligned}
\end{align}
where $\vec{v}^{(\tau_t)}(\cdot)$, $t = 1, \dots, T$, are neural networks that transform data into $p$-dimensional vectors (these neural networks are parameterised, but we do not make this explicit for notational clarity), and $g(\cdot)$ is a non-negative function (e.g., exponential or softplus) applied elementwise to its arguments. In our context of making inference from irregular spatial data, the neural networks in \eqref{eqn:quantileneuralnetwork} have architectures of the form \eqref{eqn:theta_single_field} when $\vec{Z}$ contains a single replicate, or \eqref{eqn:DeepSets} when $\vec{Z}$ contains multiple replicates, and they are also functions of the spatial locations $S$. Note that additional constraints on the parameters in $\vec{\theta}$, such as positivity, can be enforced by composing each expression in \eqref{eqn:quantileneuralnetwork} with an appropriate monotonic activation function. The neural networks in \eqref{eqn:quantileneuralnetwork} are then trained jointly by performing the optimisation task \eqref{eqn:optimisation_task} under the additive loss function,
 \begin{equation}\label{eqn:UQloss}
L(\vec{\theta}, \vec{q}^{(\tau_1)}, \dots, \vec{q}^{(\tau_T)})
=
\sum_{t = 1}^T \sum_{k=1}^p L_{\tau_t}(\theta_k, q^{(\tau_t)}_k),
 \end{equation}
where $L_\tau(\cdot, \cdot)$ is the quantile loss function given in \eqref{eqn:quantileloss_univariate} and $q^{(\tau_t)}_k$ is the $k$th element of $\vec{q}^{(\tau_t)}$. Once trained, $\vec{q}^{(\tau)}(\vec{Z})$ approximates the marginal posterior $\tau$-quantiles for $\tau \in \{\tau_1, \dots, \tau_T\}$. By including both low and high probability levels, one may construct credible intervals which, by construction, are guaranteed to be valid (i.e., non-crossing).

\section{Simulation studies}\label{sec:simulationstudies}

We now conduct several simulation studies to demonstrate the efficacy of GNN-based neural Bayes estimators for spatial models. In Section~\ref{sec:SimulationIntro}, we outline the general setting. In Section~\ref{sec:GP}, we estimate the parameters of a Gaussian process model. Since the likelihood function is available for this model, we compare our proposed estimator to the maximum-a-posteriori (MAP) estimator. In Section~\ref{sec:Schlather}, we consider a spatial extremes setting and estimate the parameters of Schlather's max-stable model \citep{Schlather_2002_max-stable_models}; the likelihood function is computationally intractable for this model, and we are able to obtain substantial improvements over the composite-likelihood approach that is often used with this model. 

\subsection{General setting}\label{sec:SimulationIntro}
 
Across the simulation studies we take the spatial domain to be the unit square. We implement our neural Bayes estimators using functionality we have added to the package \pkg{NeuralEstimators} \citep{NeuralEstimators}, which is available in the \proglang{Julia} and \proglang{R} programming languages. The GNN functionality of the package employs the \proglang{Julia} package \pkg{GraphNeuralNetworks} \citep{Lucibello2021GNN}. We conduct our experiments using a workstation with an AMD EPYC 7402 3.00GHz CPU with 52 cores and 128 GB of CPU RAM, and an Nvidia Quadro RTX 6000 GPU with 24 GB of GPU RAM. All results presented in the remainder of this paper can be generated using the reproducible code available at \url{https://github.com/msainsburydale/NeuralEstimatorsGNN}.

Our GNN architecture is based on the representation \eqref{eqn:DeepSets}. We use a propagation module based on \eqref{eqn:propagation1}--\eqref{eqn:propagation3} with $H=2$ layers and 20-dimensional hidden-feature vectors, and we define the neighbours of a node by deterministically selecting $k=30$ neighbours within a disc of fixed radius $r=0.15$ (see Section~\reffsupp{sec:neighbourhoods} for our specific selection method) where, recall, we define our domain to be the unit square $\mathcal{D} \equiv [0, 1] \times [0, 1]$.  
For the spatial-weight functions in \eqref{eqn:propagation3}, we use a combination of 10 Gaussian kernels that span the radius of the neighbourhood disc, along with an MLP with 10 output neurons, together yielding a 20-dimensional spatial-weight function for each layer. Specifically, we set
$$\vec{w}^{(l)}(d_{jj'}) = \left( \exp\Big(-\frac{(d_{jj'} - \mu_1)^2}{2\sigma_1^2}\Big), \dots, \exp\Big(-\frac{(d_{jj'} - \mu_{10})^2}{2\sigma_{10}^2}\Big), \vec{\omega}^{(l)}(d_{jj'})' \right)',$$ 
where $d_{jj'} \equiv \|\vec{s}_j - \vec{s}_{j'}\|$ denotes the Euclidean distance between spatial locations $\vec{s}_j$ and $\vec{s}_{j'}$, the means \(\mu_1, \dots, \mu_{10}\) are chosen to be the midpoints of the intervals \((0, 0.015], (0.015, 0.03], \dots, (0.135, 0.15]\), 
the standard deviations \(\sigma_1, \dots, \sigma_{10}\) are each set to 0.00375 so that each Gaussian kernel places approximately 95\% its mass within the corresponding interval, and $\vec{\omega}^{(l)}(\cdot)$ denotes an MLP with a single hidden layer of width 128. We use the elementwise average for each element of $\vec{r}(\cdot)$ in \eqref{eqn:readout} and each element of $\vec{a}(\cdot)$ in \eqref{eqn:DeepSets}. For $\vec{\phi}(\cdot)$ in \eqref{eqn:DeepSets}, we use an MLP with 128 neurons in the first two layers and $p$ neurons in the output layer, where $p$ denotes the number of parameters in the statistical model. For the final layer of $\vec{\phi}(\cdot)$, we use an exponential activation function for positive parameters and an identity activation function otherwise; for all other layers of our architecture (including those of the propagation module), we use a rectified linear unit (ReLU) activation function. In total, there are 23556 + 129$p$ neural-network parameters. We perform uncertainty quantification by jointly approximating the marginal posterior 0.025- and 0.975-quantiles, from which 95\% central credible intervals for each parameter can be constructed; our quantile network is of the form \eqref{eqn:quantileneuralnetwork}, with each $\vec{v}^{(\tau)}(\cdot)$ given by the GNN architecture described above, but with a suitable activation function for the final layer. 
 
We assume that the parameters are independent a priori and uniformly distributed on parameter-dependent intervals, ${\textrm{supp}(\theta_k)}$, $k = 1, \dots, p$. We train our neural point estimator under the mean-absolute-error loss, $L(\vec{\theta}, \hat{\vec{\theta}}) = p^{-1} \sum_{k=1}^p|\hat{\theta}_k - \theta_k|,$ so that it targets the marginal posterior medians (see Appendix~\ref{app:Proof:separablelosses}). We set $K$ in \eqref{eqn:optimisation_task} to 10000 and 2000 for the training and validation parameter sets, respectively. Simulation from the statistical models that we consider requires matrix factorisation for each parameter-vector and spatial-configuration pair. To reduce training time, we therefore keep the training and validation parameter sets fixed, and construct the training and validation data by simulating multiple (specifically, five) data sets for each parameter vector in the training and validation sets \citep[as was done by, e.g.,][]{Gerber_Nychka_2021_NN_param_estimation, Sainsbury-Dale_2022_neural_Bayes_estimators}. Our training and validation data sets are simulated using spatial configurations, $S$, sampled from a Matérn cluster process on the spatial domain \mbox{$\mathcal{D}$} and whose parameters vary uniformly between the values illustrated in Figure~\ref{fig:spatialpatterns} (with the expected number of sampled points in each field, $\lambda\mu$, fixed to $250$). During training, we simulate the training data ``on-the-fly'' to reduce overfitting \citep[see][Sec.~2.3]{Sainsbury-Dale_2022_neural_Bayes_estimators}. We cease training when the empirical risk in \eqref{eqn:optimisation_task} evaluated using the validation set has not decreased in five consecutive epochs. 
 
We compare the trained neural point estimator to likelihood-based estimators using several synthetic data sets with spatial configurations that are unlikely to occur as realisations of the cluster process used to sample $S$ during training, in order to assess the robustness of the neural point estimator to unexpected configurations $S$. To assess our neural credible intervals, we empirically estimate a marginal version of the expected coverage \citep[][Definition~2.1]{Hermans_2022} and compare it to the nominal expected coverage level. 

\subsection{Gaussian process model}\label{sec:GP}

In this subsection, we consider a classic spatial model, the Gaussian process model, with a single spatial replicate (i.e., $m = 1$). The data model is
\begin{equation}\label{eqn:GP}
Z_{j} = Y(\vec{s}_{j}) + \epsilon_{j}, \; j = 1, \dots, n,
\end{equation}
 where $\vec{Z} \equiv (Z_{1}, \dots, Z_{n})'$ are data observed at locations \mbox{$\{\vec{s}_{1}, \dots, \vec{s}_{n}\} \subset \mathcal{D}$}, $Y(\cdot)$ is a spatially-correlated mean-zero Gaussian process, and \mbox{$\epsilon_{j} \sim \Gau(0, \sigma_\epsilon^2)$}, $j = 1, \dots, n$. Spatial dependence is captured through the covariance function, $C(\vec{s}, \vec{u}) \equiv \cov{Y(\vec{s})}{Y(\vec{u})}$, for $\vec{s}, \vec{u} \in \mathcal{D}$. Here, we use the popular isotropic Mat\'{e}rn covariance function, 
\begin{equation}\label{eqn:Matern_covariance_function}
C(\vec{s}, \vec{u}) = \sigma^2 \frac{2^{1 - \nu}}{\Gamma(\nu)} \Big(\frac{\|\vec{s} - \vec{u}\|}{\rho}\Big)^\nu K_\nu\Big(\frac{\|\vec{s} - \vec{u}\|}{\rho}\Big), 
\end{equation} 
where $\sigma^2$ is the marginal variance, $\Gamma(\cdot)$ is the gamma function, $K_\nu(\cdot)$ is the Bessel function of the second kind of order $\nu$, and $\rho > 0$ and $\nu > 0$ are range and smoothness parameters, respectively. For ease of illustration, we fix $\sigma^2 = 1$ and $\nu = 1$, which leaves two unknown parameters that need to be estimated: $\vec{\theta} \equiv (\sigma_\epsilon, \rho)'$. In Section~\ref{sec:application}, we illustrate a case where we also estimate $\sigma^2$.     

\begin{figure*}[t!]
    \centering
    \includegraphics[width = \textwidth]{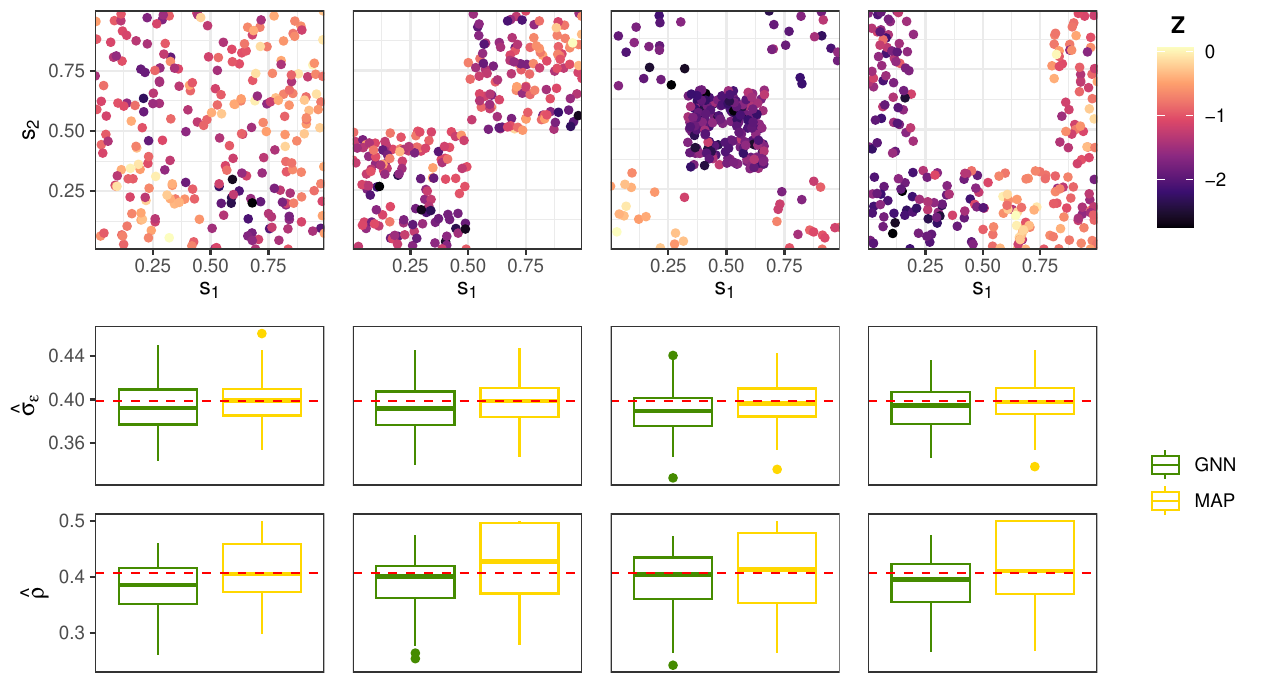}  
    \caption{Several spatial data sets (top row) and empirical marginal sampling distributions (second and third rows) of two estimators for the Gaussian process model of Section~\ref{sec:GP} with true parameters denoted by the dashed line. The estimators are a GNN-based neural Bayes estimator and the maximum a posteriori (MAP) estimator. A single GNN was trained for all data sets. 
    }\label{fig:GP}
\end{figure*}

We use the priors \mbox{$\sigma_\epsilon \sim \Unif{0}{1}$} and \mbox{$\rho \sim \Unif{0.05}{0.5}$}. The total training time for our GNN-based estimator is 24 minutes. In our implementation, the MAP estimator takes 1.2 seconds to estimate the parameters from a single data set with $n = 250$ spatial locations, while the GNN-based estimator takes 0.002 seconds; a 600-fold speedup post training. Figure~\ref{fig:GP} shows the empirical sampling distributions of both our GNN-based estimator and the MAP estimator under a single parameter configuration, but over four different spatial configurations (which were not in the set of locations used to train the GNN-based estimator), all with $n=250$ locations. Although our neural Bayes estimator and the MAP estimator are associated with different loss functions, both estimators are approximately unbiased and have similar variances. Next, to quantify the overall performance of the estimators, we construct a test set of 1000 parameter vectors sampled from the prior distribution and for each parameter vector, a data set for each spatial configuration shown in Figure~\ref{fig:GP}, yielding a total of 4000 data sets. We then compute the empirical root-mean-squared error (RMSE) for each each estimator from these data sets. The RMSE values for the GNN-based and MAP estimator are 0.050 and 0.046, respectively. Our GNN-based estimator therefore performs nearly as well as the MAP estimator in terms of RMSE, and it is clearly able to make inference from a wide range of spatial configurations.

Having established the efficacy of GNN-based point estimation, we next consider uncertainty quantification. Following the methodology described in Section~\ref{sec:UQ}, we construct a neural Bayes estimator that approximates the 0.025 and 0.975 marginal posterior quantiles, and use these to construct credible intervals with 95\% nominal expected coverage. The training time is 52 minutes, while estimation from a single data set with $n=250$ locations takes 0.004 seconds. We assess these intervals by sampling 3000 parameter vectors from the prior distribution and, for each parameter vector, a set of spatial locations sampled from the previously described Matérn cluster process; simulating 10 data sets for each parameter-vector and spatial-configuration pair; and computing the overall empirical coverage from these 30000 data sets. The empirical coverages for $\rho$ and $\sigma_\epsilon$ are 95.2\% and 94.6\%, respectively, which are close to the nominal value. 

These results show that our GNN-based neural Bayes estimator is performing as one would expect and that it can be applied to data sets with differing spatial configurations. For the Gaussian process model, inference with the likelihood function is feasible and neural Bayes estimators are usually not required unless one needs to do estimation repeatedly, as we illustrate in Section~\ref{sec:application}. Neural Bayes estimators are particularly beneficial when the likelihood function is unavailable, as is the case for the model we consider next. 

\subsection{Schlather's max-stable model}\label{sec:Schlather}

Despite their limitations \citep{Huser_2024_max-stable}, max-stable processes remain a central pillar of spatial extreme-value analysis \citep{Davison_Huser_2015_Statistics_of_extremes, Davison_2019_spatial_extremes, Huser_2022_advances_in_spatial_extremes}, being the only possible non-degenerate limits of properly renormalised pointwise block maxima of independent and identically distributed (i.i.d.) random fields. However, inference using the full likelihood function is computationally infeasible with even a moderate number of observed locations \citep{Castruccio_2016_composite_likelihood_max-stable}; they are, therefore, ideal candidates for likelihood-free inference. Here we consider Schlather's max-stable model \citep{Schlather_2002_max-stable_models}, given by
\begin{equation}\label{eqn:SchlathersModel}
Z_{ij} = 
\max_{k \in \mathbb{N}}
\zeta_{ik}^{-1} \max\{0, Y_{ik}(\vec{s}_{ij})\}, \quad  i = 1, \dots, m, \; j = 1, \dots, n_i, 
\end{equation}
where, for replicates $i = 1, \dots, m$, $\vec{Z}_i \equiv (Z_{i1}, \dots, Z_{in_i})'$ are observed at locations \mbox{$\{\vec{s}_{i1}, \dots, \vec{s}_{in_i}\} \subset \mathcal{D}$}, $\{\zeta_{ik} : k \in \mathbb{N}\}$ are i.i.d.~Poisson point processes on $(0, \infty)$ with unit intensity, and \mbox{$\{Y_{ik}(\cdot) : k \in \mathbb{N}\}$} are i.i.d.~mean-zero Gaussian processes scaled so that $\E[{\max\{0, Y_{ik}(\cdot)\}}] = 1$. Here, we model each $Y_{ik}(\cdot)$ using the Mat\'{e}rn covariance function \eqref{eqn:Matern_covariance_function}, with $\sigma^2 = 1$. Hence, the unknown parameter vector to estimate is $\vec{\theta} \equiv (\rho, \nu)'$. 

We compare our GNN-based estimator to a likelihood-based estimator; however, for max-stable models, the likelihood function is computationally intractable, since the number of terms grows super-exponentially fast in the number of observed locations \citep[see, e.g.,][]{Padoan_2010_composite_likelihood_max_stable_processes, Huser_2019_advances_in_spatial_extremes}. A popular substitute is the pairwise likelihood (PL) function, a composite likelihood formed by considering only pairs of observed locations. Specifically, the pairwise log-likelihood function for the $i$th replicate is
  \begin{equation}\label{eqn:pairwise_likelihood}
 \ell_{\rm{PL}}(\vec{\theta}; \vec{Z}_i) \equiv \sum_{j = 1}^{n_i - 1}\sum_{j' = j + 1}^{n_i}  \omega^{(i)}_{jj'}\log f(Z_{ij}, Z_{ij'} \mid \vec{\theta}), 
 \end{equation} 
where $f(\cdot, \cdot \mid \vec{\theta})$ denotes the bivariate probability density function for pairs in $\vec{Z}_i$ \citep[see][pg.~231--232]{Huser_2013_PhD_thesis} and $\omega^{(i)}_{jj'}$ denotes a nonnegative weight. 
 Hence, here we compare our GNN-based estimator to the pairwise MAP (PMAP) estimator, 
\begin{equation*}
\hat{\vec{\theta}}_{\rm{PMAP}}(\vec{Z})
=
\argmax_{\vec{\theta}}
\sum_{i=1}^m \ell_{\rm{PL}}(\vec{\theta}; \vec{Z}_i) + 
\log{\pi(\vec{\theta})},
\end{equation*}
where $\pi(\vec{\theta})$ denotes the prior density function. 
Note that, in contrast to the more commonly used PL estimator, the PMAP estimator incorporates prior information, which facilitates a fair comparison to our neural Bayes estimator. 
The computational and statistical efficiency of the PMAP estimator can often be improved by constructing \eqref{eqn:pairwise_likelihood} using only a subset of pairs that are within a fixed cut-off distance \citep{Bevilacqua_2012, Sang_Genton_2014}; here, we find that considering pairs within a distance of 0.2 units provides the best results and, therefore, we set $\omega^{(i)}_{jj'} = \mathbb{I}(\|\vec{s}_{ij}-\vec{s}_{ij'}\|\leq 0.2)$ in \eqref{eqn:pairwise_likelihood}.   

\begin{figure*}[t!]
    \centering
    \includegraphics[width = \textwidth]{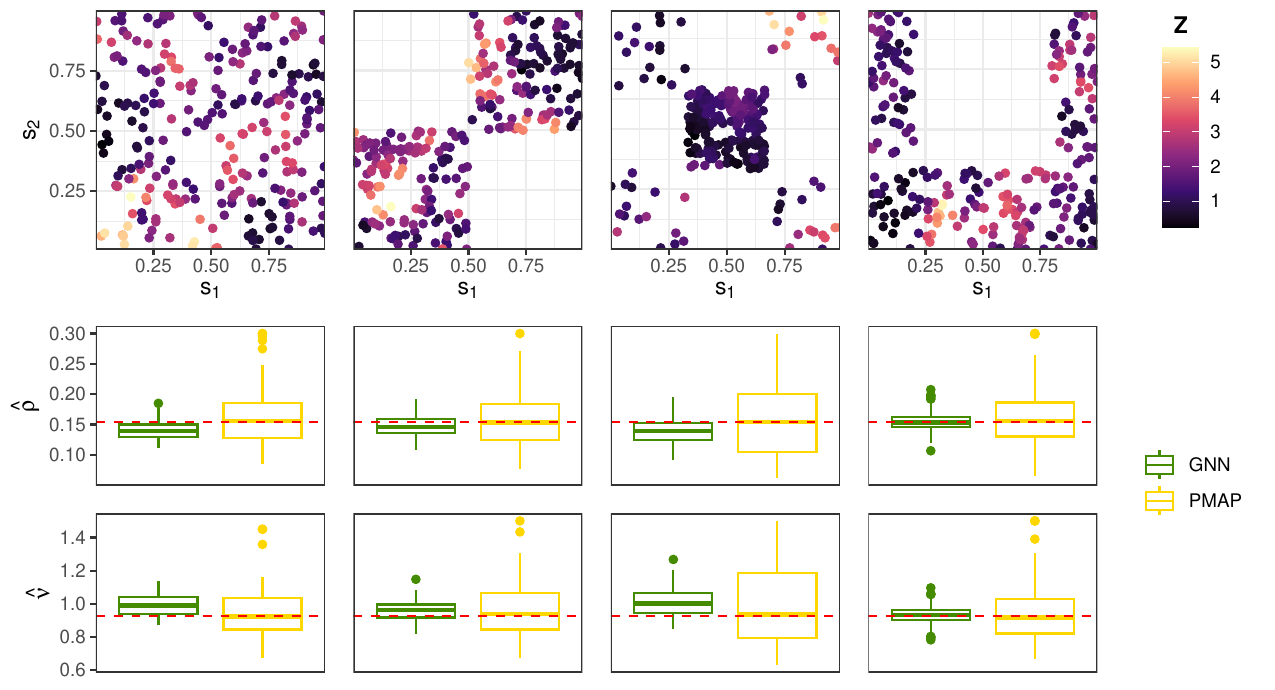}  
    \caption{Several spatial data sets (top row) and corresponding empirical marginal sampling distributions (second and third rows) of two estimators for Schlather's max-stable model of Section~\ref{sec:Schlather} with true parameters denoted by the dashed line. The estimators are a GNN-based neural Bayes estimator and the pairwise maximum a posteriori (PMAP) estimator. 
    }\label{fig:Schlather}
\end{figure*}

We use the priors \mbox{$\rho \sim \Unif{0.05}{0.3}$} and \mbox{$\nu \sim \Unif{0.5}{2.5}$}, and we consider $m = 20$ independent spatial fields for each parameter vector, with locations sampled during training according to the Matérn cluster process with $n=250$ locations on average, as in Section~\ref{sec:SimulationIntro}. Realisations from the present model, here expressed on unit Fréchet margins, tend to have highly varying magnitudes. We reduce this variability by log-transforming our data to the unit Gumbel scale. The total training time for our GNN-based estimator is 53 minutes. The PMAP estimator takes about 11.5 seconds to estimate the parameters from a single data set, while the GNN-based estimator takes 0.002 seconds, a 5750-fold speedup post training. Figure~\ref{fig:Schlather} shows the empirical sampling distributions of both our GNN-based estimator and the PMAP estimator under a single parameter configuration but over four different spatial sample configurations. Both estimators are approximately unbiased, but the GNN-based estimator has lower variance. Next, to quantify the overall performance of the estimators, we construct a test set of 4000 data sets as detailed in Section~\ref{sec:GP}, and compute the empirical RMSE for both estimators. The RMSE of the GNN-based and PMAP estimator is 0.056 and 0.126, respectively; our proposed estimator therefore provides a substantial improvement over the PMAP estimator. 

We next consider uncertainty quantification performed with a neural Bayes estimator that approximates the marginal posterior 0.025- and 0.975-quantiles. The training time is 2.87 hours, while estimation from a single data set with $n=250$ locations takes 0.004 seconds. As in Section~\ref{sec:GP}, we assess the accuracy of the credible intervals using the overall empirical coverage from 30000 simulated data sets, with the spatial locations sampled from the Matérn cluster process described above. The empirical coverages for $\rho$ and $\nu$ are 95.7\% and 96.3\%, respectively, which are close to the nominal value.
 
 Overall, we find that for Schlather’s max-stable model, the proposed GNN-based neural Bayes estimator is superior to the estimator based on the pairwise likelihood function.

\section{Application to global sea-surface temperature data}\label{sec:application}

\begin{figure}
  \includegraphics[width = \textwidth]{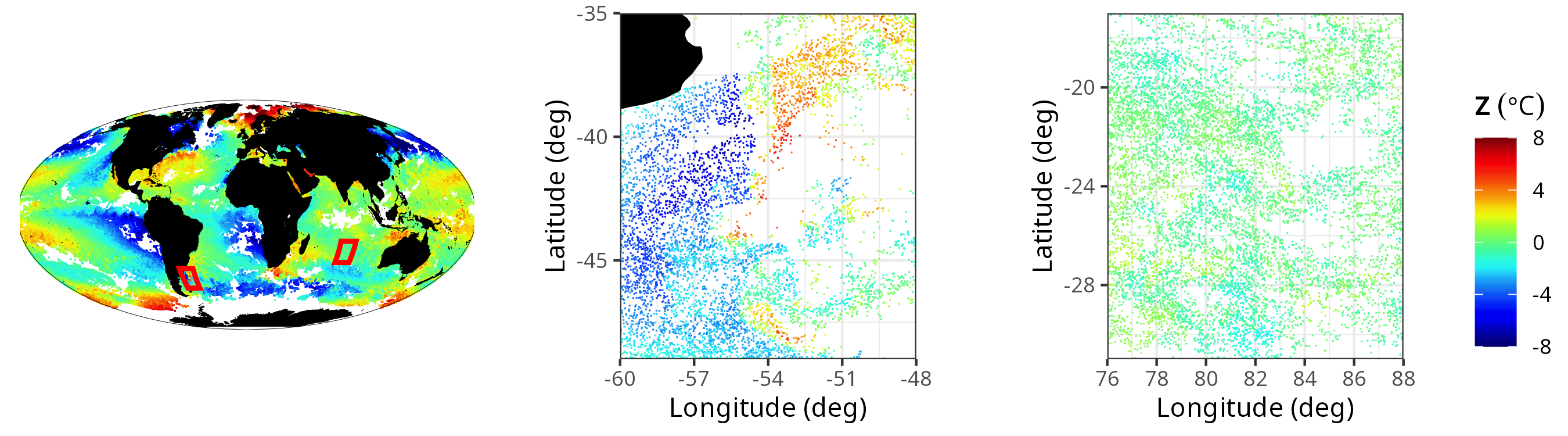}
  \caption{  
 SST residuals over the globe (left), in the Brazil-Malvinas Confluence Zone (centre), and in the southern Indian Ocean (right). These regions, which are demarcated by rectangles in the left panel, illustrate the spatial nonstationarity present in this data set. 
  }\label{fig:SST:data}
\end{figure}

We now apply our methodology to the analysis of a massive global sea-surface temperature (SST) data set. Our application uses the data analysed by \cite{Zammit_2020} and \cite{Cressie_2021_review_spatial-basis-function_models}, which consists of SST data obtained from the Visible Infrared Imaging Radiometer Suite (VIIRS) on board the Suomi National Polar-orbiting Partnership (Suomi NPP) weather satellite \citep{Cao_2013_Suomi_NPP}. The data set consists of one million observations. As in \cite{Zammit_2020}, we model the spatial residuals from a linear model with covariates given by an intercept, the latitude coordinate, and the square of the latitude coordinate. Figure~\ref{fig:SST:data} shows these detrended data over the globe and in two regions corresponding to the Brazil-Malvinas Confluence Zone and the southern Indian Ocean. There is clear evidence of spatial covariance nonstationarity. 

 To account for nonstationarity, we take a local modelling approach by partitioning the spatial domain and fitting a separate model within each region. Our partitioning is the ISEA Aperture 3 Hexagon (ISEA3H) discrete global grid (DGG) at resolution~5, which contains 2432 equally-sized hexagonal cells. We model the dependence structure within each hexagon using the Gaussian process model of Section~\ref{sec:GP}, with unknown range parameter, $\rho$, process standard deviation, $\sigma$, and measurement-error standard deviation, $\sigma_\epsilon$. Therefore, within each cell, we estimate three parameters, $\vec{\theta} \equiv (\rho, \sigma, \sigma_\epsilon)'$. We adopt a moving-window approach \citep{Haas_1990_moving_window, Haas_1990_moving_window_lognormal, Kuusela_Stein_2018, Castro-Camilo_2020_local_estimation} to parameter estimation, whereby the parameter estimates for a given cell are obtained using both the data within that cell and the data within its neighbouring cells. We refer to a cell and its neighbours as a cell cluster; the left panel of Figure~\reffsupp{fig:SST:clustering} shows an example of two cell clusters. This moving-window approach makes large-scale trends more apparent and allows us to obtain estimates in unobserved cells, provided that neighbouring cells contain data. In total, there are 2161 cell clusters that contain data; these clusters contain a median number of 2769 observed locations, and a maximum of 12591 observed locations; the right panel of Figure~\reff{fig:SST:clustering} shows a histogram of the number of observed locations for all cell clusters.

For point estimation we use a single GNN-based neural Bayes estimator trained under the mean-absolute-error loss. For uncertainty quantification we obtain credible intervals by approximating the marginal posterior 0.025- and 0.975-quantiles jointly using a single GNN-based neural Bayes estimator, as described in Section~\ref{sec:UQ}. We use the same architectures described in Section~\ref{sec:SimulationIntro}. Since the amount of available data varies between cell clusters, we train our estimator using simulated spatial data sets with variable sample size; each set of spatial locations, $S$, used to construct the training data, is sampled from a Matérn cluster process (recall Figure~\ref{fig:spatialpatterns}) on the unit square, with the expected number of sampled points varying between $n = 30$ and $n = 2000$. To estimate the parameters in cell clusters with a higher number of observed locations, we make use of the estimator’s ability to extrapolate to values of $n$ larger than those used during training  (a property illustrated and discussed in Section~\reffsupp{sec:experiment:variablesamplesizes}). Note that we could train our estimator with a distribution on $n$ based on the distribution of sample sizes in our data set, shown in Figure~\reff{fig:SST:clustering}; however, we choose not to do so since we prefer to illustrate the use of a single, broadly-applicable GNN-based neural Bayes estimator, rather than one tailored specifically to this data set. Since we train our estimator using spatial locations sampled within the unit square, our estimator is calibrated for distances in $[0, \sqrt{2}]$. Therefore, as a pre-processing step, we scale the (chordal) distances within each cell cluster to be within this range; the estimated range parameter is then scaled back for interpretation. The use of chordal distance is justified by the small size of the cells: it is reasonable to model the Earth's surface as flat within the cell clusters. In this application, since we train our neural networks once and apply the resulting point and quantile estimators to data with widely varying dependence structures, it is important that a vague prior is used. Here, we assume that our parameters are independent a priori with marginal priors \mbox{$\rho \sim \Unif{0.05}{0.60}$}, \mbox{$\sigma \sim \Unif{0}{3}$}, and \mbox{$\sigma_\epsilon \sim \Unif{0}{1}$}. The total training time is about 4 hours. We assess our trained estimators using the simulation-based (empirical) approach from Section~\ref{sec:simulationstudies}; see Figure~\reffsupp{fig:GPSigmaVaried}. Our neural credible intervals for $\rho$, $\sigma$, and $\sigma_\epsilon$ were found to have empirical coverages of 95.2\%, 94.1\%, and 95.1\%, respectively, which are close to the nominal value of 95\%. 
 
\begin{figure}
  \includegraphics[width = \textwidth]{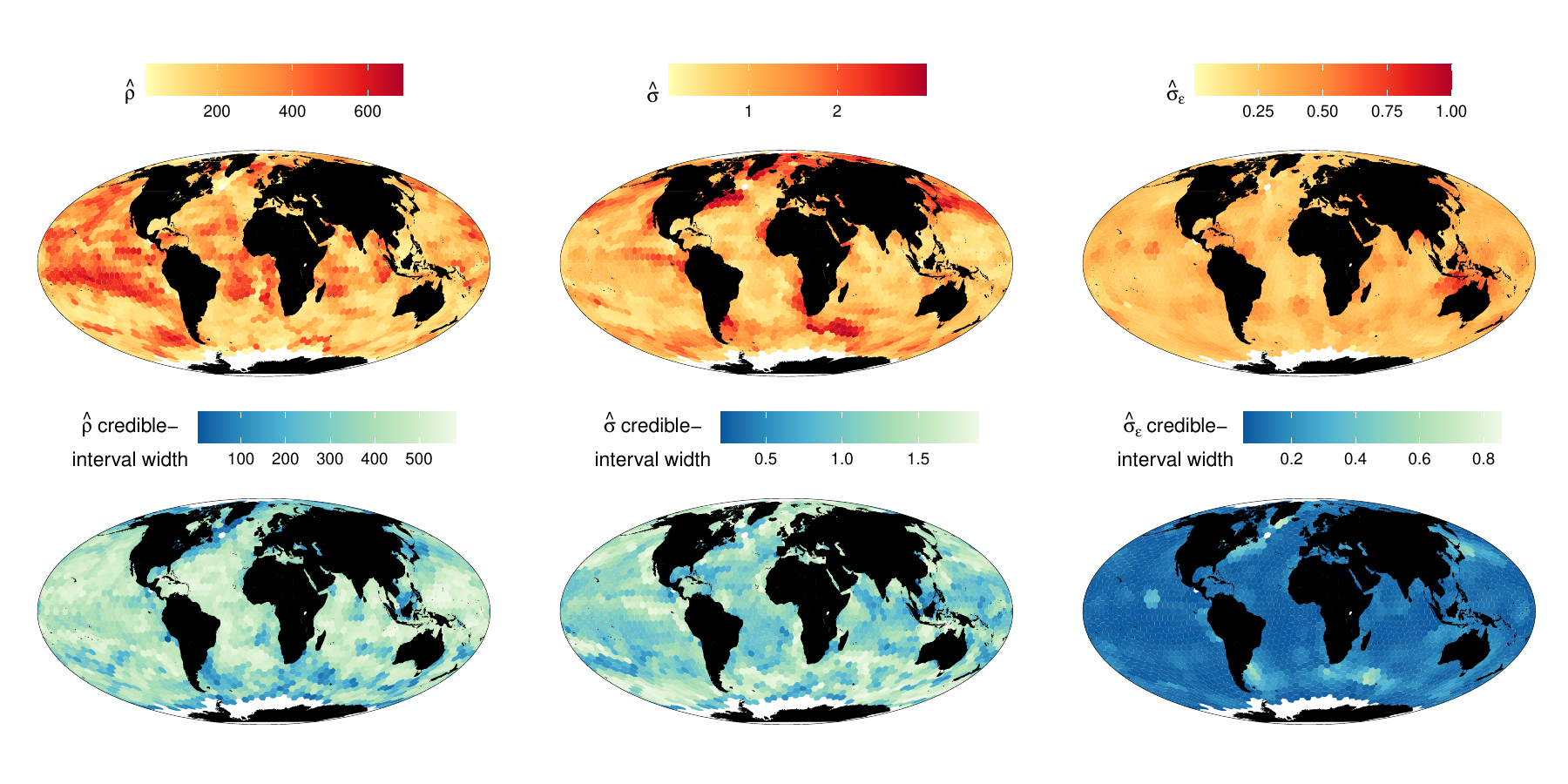}
  \caption{  
Spatially varying point estimates (top row) and corresponding 95\% credible-interval widths (bottom row) for each parameter of the Gaussian process model used to analyse global SST in Section~\ref{sec:application}. The first, second, and third columns correspond to the range parameter, $\rho$, process standard deviation, $\sigma$, and measurement-error standard deviation, $\sigma_\epsilon$. The globe is partitioned using the ISEA Aperture 3 Hexagon (ISEA3H) discrete global grid (DGG) at resolution 5. 
  }\label{fig:SST:estimates}
\end{figure}

Figure~\ref{fig:SST:estimates} shows spatially varying point estimates and 95\% credible-interval widths for each parameter. Figure~\reffsupp{fig:SST:intervals} shows estimates of the 0.025- and 0.975-quantiles. Our neural Bayes estimators provide point estimates and credible intervals over 2161 cell clusters in just over three minutes. The point estimates given in Figure~\ref{fig:SST:estimates} conform with what one may expect when modelling global SST: energetic regions, for example, near the South-East coast of South America, tend to exhibit large estimates of the process standard deviation, $\sigma$, and small estimates of the length scale $\rho$; by contrast, more stable regions, such as those towards the centre of large ocean basins, tend to exhibit small estimates of $\sigma$ and larger estimates of $\rho$. 

Due to the scale of the estimation task, it is computationally prohibitive to validate our point estimates and credible intervals using asymptotically exact methods such as Markov chain Monte Carlo (MCMC). We therefore only validate our point estimates, by comparing our neural point estimates to MAP estimates. To ease the computational burden, we cap the number of spatial data in each cell cluster to $n=3000$ when computing MAP estimates; even then, MAP estimation with this restricted data set takes slightly over 10 hours. Figure~\reffsupp{fig:SST:GNN_ML_estimates} compares estimates from the neural Bayes estimator to those from the MAP estimator. There is some discrepancy between the estimates of the length scale $\rho$; this could be due to the fact that the MAP estimates are based on a maximum of 3000 data points per region. Estimates of the remaining two parameters are mostly in agreement. Finally, we note that conventional goodness-of-fit tests may also be used when a model is fit with a neural Bayes estimator; however, assessing the appropriateness of the Gaussian process model for this particular SST data set is beyond the scope of the paper. 

\section{Conclusion}\label{sec:conclusion}

In this paper, we develop a new approach to neural Bayes estimation from irregular spatial data that uses GNNs. Our approach has two main strengths. First, GNN-based neural Bayes estimators are specifically designed to capture spatial dependence, and are thus parsimonious approximators of Bayes estimators in spatial settings. Second, GNN-based estimators can be applied to data collected over any set of spatial locations, which allows the computationally-intensive training step to be amortised for a given spatial model. That is, a single GNN-based estimator can be re-used for new spatial data sets irrespective of the new observation locations. Importantly, we also combine the GNN architecture with the DeepSets framework to construct a neural Bayes estimator applicable to any number of independent replicates, thus opening the door to amortised estimation in a wide range of application settings for arbitrary spatial models. We provide implementation guidelines pertaining to neural-network-architecture design and the construction of synthetic spatial data sets for training the estimators. We also perform uncertainty quantification via a suitably designed neural Bayes estimator that approximates a set of marginal posterior quantiles (and that avoids quantile crossing). Finally, we provide user-friendly access to our methodology by incorporating it within the package \pkg{NeuralEstimators} \citep{NeuralEstimators}, which is available in the \proglang{Julia} and \proglang{R} programming languages. 

The extension to irregular spatio-temporal data using spatio-temporal GNNs \citep[Sec.~VII]{Wu_2021_GNN_review} is the subject of future work. GNNs also extend naturally to multivariate spatial processes \citep{Gneiting_2010_multivariate_Matern,  Genton_2015_multivariate_geostatistics, Genton_2015_multivariate_maxstable_processes}, although the often complicated parameter constraints in these settings require careful consideration. Our architecture is tailored to isotropic spatial dependence models; more general architectures \citep[e.g.,][]{Danel_2020_spatial_GNN} may be needed for other models, for example those exhibiting strong nonstationarity or anisotropy. We have focused on point estimation; GNNs, however, would also be useful for approximating the likelihood function or the full posterior distribution of spatial-model parameters, for instance by incorporating them as a module in a normalising flow \citep[e.g.,][]{Radev_2022_BayesFlow}, as was done in the context of agent-based modelling by \cite{Dyer_2022_GNN}. Alternatively, GNNs could be used to automatically learn relevant summary statistics for use in approximate Bayesian computation \citep[ABC; see, e.g.,][]{Jiang_2017, Chen_2021_NN-approximate_sufficient_statistics_for_implicit_models}, which can also be used for amortised inference \citep{Mestdagh_2019_prepaid_ABC}. It is also straightforward to combine GNNs with the censoring framework of \citet{Richards_2023_censoring}, in order to perform inference from censored data collected over arbitrary spatial locations. GNNs may also prove useful in non-spatial applications; for example, exponential random graph models \citep[ERGMs;][]{Robins_2007_ERGMs, Lusher_2013_ERGMs} used in network analysis have a normalising constant that prevents straightforward evaluation of the likelihood function, and would therefore benefit from the proposed likelihood-free methodology. Future research will compare risk-minimisation approaches (e.g., neural Bayes estimation) to conventional sampling-based likelihood-free methods (e.g., ABC), particularly with respect to the number of model simulations required to make accurate inferences in the tails. Finally, GNNs could complement existing likelihood-based approaches, for example by providing good initial estimates for maximum-likelihood estimation, and such a ``semi-amortised'' approach \citep{Hjelm_2016} could lead to reduced run-times of classical optimisation-based estimation algorithms. 



\section*{Acknowledgements}

The authors would like to thank Noel Cressie for discussion and feedback. We are also grateful to the reviewers and the editors for their helpful comments and suggestions that improved the quality of the manuscript. 

\section*{Funding}

Matthew Sainsbury-Dale's and Andrew Zammit-Mangion’s research was supported by an Australian Research Council Discovery Early Career Research Award, DE180100203. Matthew Sainsbury-Dale’s research was further supported by an Australian Government Research Training Program Scholarship, a 2022 Statistical Society of Australia top-up scholarship, the King Abdullah University of Science and Technology (KAUST) Office of Sponsored Research (OSR) Award No. OSR-CRG2020-4394, and Rapha\"el Huser's baseline funds. Jordan Richard's and Raphaël Huser's research was also supported by KAUST OSR-CRG2020-4394 and Rapha\"el Huser's baseline funds. This publication is based upon work supported by KAUST Research Funding (KRF) under Award No. ORFS-2023-OFP-5550, and on work supported by the Air Force Office of Scientific Research under award number FA2386-23-1-4100. Andrew Zammit-Mangion also acknowledges the Swiss National Science Foundation for travel support to present this research (grant no. IZSEZ0\_225139). 


 
\appendix 
\section{Invariance of the Bayes estimator under different point process distributions}\label{app:Proof:priormeasureS}

In this appendix we show that when the spatial locations of the data are treated as a realisation of a point process, the Bayes estimator is, under certain conditions, invariant to the distribution of the point process. For ease of exposition, we consider the case where the posterior distribution admits a density function with respect to Lebesgue measure. Furthermore, while formally point process realisations are locally finite counting measures, we here view them through their associated point patterns; thus, for a point process $S$ and a space of locally finite configurations $\mathcal{S}$, we write $S\in\mathcal{S}$ as shorthand for `a realisation of $S$ with associated point pattern taking values in $\mathcal{S}$'. For generic random quantities $A$ and $B$, we use $[A \mid B]$ to denote the conditional probability density function of $A$ given $B$. 

\begin{theorem}\label{thm:GNNproof} 
 Denote by $\mathcal{S}$ the space of all locally finite point patterns on a spatial domain $D \subset \mathbb{R}^2$, and let $S$ be a point process associated with point patterns taking values in $\mathcal{S}$. Let the data $\vec{Z} \in \samplespace_S \subseteq \mathbb{R}^{|S|}$ have distribution that is conditional on $S \in \mathcal{S}$ and $\vec{\theta}  \in \Theta \subseteq \mathbb{R}^p$. Let $L: \Theta \times \Theta \to \mathbb{R}^{\geq 0}$ denote a strictly convex nonnegative loss function. Assume that the Bayes estimate $\hat{\vec{\theta}}^\star$ has finite posterior expected loss  $\int_\Theta L(\vec{\theta}, \hat{\vec{\theta}}^\star)[\vec{\theta} \mid \vec{Z}, S]\emph{\d}\vec{\theta}$ for all fixed $\vec{Z} \in \mathcal{Z}_S$ and $S \in \mathcal{S}$. Then the Bayes estimator $\hat{\vec{\theta}}^\star(\vec{Z}, S)$ is invariant to the distribution of $S$ provided 
 
 \begin{enumerate}[(i)]
     \item its induced probability measure ${\Omega}(\cdot)$ is strictly positive (i.e., has strictly positive measure on all non-empty open Borel subsets of $\mathcal{S}$), and,
     \item $S$ and $\vec{\theta}$ are independent.
\end{enumerate}
     \end{theorem}

\begin{proof}

For all fixed $\vec{Z} \in \mathcal{Z}_S$ and $S \in \mathcal{S}$, a Bayes estimate $\hat{\vec{\theta}}^\star$ minimises the posterior expected loss, that is, 
\begin{equation}
\hat{\vec{\theta}}^\star = \argmin_{\hat{\vec{\theta}}} \int_\Theta L(\vec{\theta}, \hat{\vec{\theta}})[\vec{\theta} \mid \vec{Z}, S]\d\vec{\theta}.\label{eq:Bayesestimate}
\end{equation}
By assumption, $\int_\Theta L(\vec{\theta}, \hat{\vec{\theta}}^\star)[\vec{\theta} \mid \vec{Z}, S]\d\vec{\theta} < \infty$ and, since $L(\cdot,\cdot)$ is strictly convex, the estimate is unique \citep[][Ch. 4, Cor. 1.4]{Lehmann_Casella_1998_Point_Estimation}. Consider now the Bayes estimator  $\hat{\vec{\theta}}^\star(\vec{Z}, S)$ that returns the Bayes estimate for any fixed $\vec{Z} \in \mathcal{Z}_S$ and $S \in \mathcal{S}$ \citep[see][for a proof of the existence of a Borel measurable Bayes estimator under mild conditions]{Brown_1973}. 
Since the posterior expected loss is bounded and nonnegative for all $\vec{Z} \in \mathcal{Z}_S$ and $S \in \mathcal{S}$, 
we can also assert that 
\begin{equation}\label{eq:Bayesmin}
  \hat{\vec{\theta}}^\star(\cdot,\cdot) = 
 \argmin_{\hat{\vec{\theta}}(\cdot,\cdot)}
  \int_\mathcal{S}\int_\mathcal{Z}\int_\Theta L(\vec{\theta}, \hat{\vec{\theta}}(\vec{Z},S))[\vec{\theta} \mid \vec{Z}, S]\d\vec{\theta} \d \widetilde{F}_S(\vec{Z}) \d{\Omega}(S),
  \end{equation}
for any strictly positive conditional (on $S$) probability measure $\widetilde{F}_S(\cdot)$ and any strictly positive probability measure ${\Omega}(\cdot)$. Choosing $\d \widetilde{F}_S(\vec{Z}) = [\vec{Z} \mid S]\d\vec{Z}$ for the conditional measure in \eqref{eq:Bayesmin}, we see that
$$
  \hat{\vec{\theta}}^\star(\cdot,\cdot) = \argmin_{{\vec{\theta}}(\cdot,\cdot)}\int_\mathcal{S}\int_\mathcal{Z}\int_\Theta L(\vec{\theta}, \hat{\vec{\theta}}(\vec{Z},S))[\vec{\theta} \mid \vec{Z},S]\d\vec{\theta}  [\vec{Z} \mid S]  \d\vec{Z}\d\Omega(S).
$$
Applying Bayes rule to $[\vec{\theta} \mid \vec{Z},S]$ and assuming $S$ and $\vec{\theta}$ are independent yields Equation~\eqref{eq:riskv2}, thus completing the proof.
\end{proof}

 \section{Bayes estimators under additive loss functions}\label{app:Proof:separablelosses}

In this appendix we show that a Bayes estimator with respect to a sum of univariate loss functions is given by the vector of marginal Bayes estimators. As in Appendix~\ref{app:Proof:priormeasureS}, for ease of exposition we consider the case where the posterior distribution admits a density function with respect to Lebesgue measure, and we use $[A \mid B]$ to denote the conditional probability density function of $A$ given $B$. We use $\vec{\theta}_{\setminus k}$ to denote the vector $\vec{\theta}$ with its $k$th element removed. Similarly, $\Theta_{k}$ and $\Theta_{\setminus k}$ denote the spaces of $\theta_k$ and $\vec{\theta}_{\setminus k}$, respectively. 

  \begin{theorem} \label{theorem:additive_loss_functions}
  \textit{Let the data $\vec{Z}$ be distributed according to a family of distributions indexed by $\vec{\theta}  \in \Theta \subseteq \mathbb{R}^p$ on a sample space $\samplespace$. Let $L: \Theta \times \Theta \to \mathbb{R}^{\geq 0}$ denote a loss function of the form
\begin{equation}\label{eqn:jointloss_from_marginal}
L(\vec{\theta}, \hat{\vec{\theta}}) \equiv \sum_{k=1}^p L_k(\theta_k, \hat{\theta}_k), 
\end{equation}
where, for $k = 1, \dots, p$, $L_k: \Theta_{k} \times \Theta_{k} \to \mathbb{R}^{\geq 0}$ is a univariate loss function. Then a Bayes estimator under $L(\cdot, \cdot)$ is given by $(\hat{\theta}_1^\star(\cdot), \dots, \hat{\theta}_p^\star(\cdot))'$ where, for $k = 1, \dots, p$, $\hat{\theta}_k^\star(\cdot)$ is a Bayes estimator for $\theta_k$ under the loss function $L_k(\cdot, \cdot)$. 
} 
  \end{theorem}

  \begin{proof*}
  \noindent Provided that there exists an estimator with finite Bayes risk, a Bayes estimator for any given $\vec{Z} \in \samplespace$ can be obtained by minimising the posterior expected loss (see, e.g., \citeauthor{Lehmann_Casella_1998_Point_Estimation}, \citeyear{{Lehmann_Casella_1998_Point_Estimation}}, Ch.~4, Thm.~1.1; \citeauthor{Robert_2007_The_Bayesian_Choice}, \citeyear{Robert_2007_The_Bayesian_Choice}, Thm.~2.3.2), 
  \begin{equation}\label{eqn:conditional_risk}
  \int_\Theta L(\vec{\theta}, \hat{\vec{\theta}}(\vec{Z}))[\vec{\theta} \mid \vec{Z}] \d\vec{\theta}.
  \end{equation}
Under the loss function \eqref{eqn:jointloss_from_marginal}, the posterior expected loss \eqref{eqn:conditional_risk} is
\begin{align*}
\int_\Theta L(\vec{\theta}, \hat{\vec{\theta}}(\vec{Z}))[\vec{\theta} \mid \vec{Z}] \d\vec{\theta}
&= \int_\Theta \sum_{k=1}^p L_k(\theta_k, \hat{\theta}_k(\vec{Z}))  [\vec{\theta} \mid \vec{Z}] \d\vec{\theta} \\
&= \sum_{k=1}^p \int_\Theta L_k(\theta_k, \hat{\theta}_k(\vec{Z}))  [\theta_k \mid \vec{Z}][\vec{\theta}_{\setminus k} \mid \theta_k, \vec{Z}] \d\vec{\theta}\\
&= \sum_{k=1}^p \int_{\Theta_k} L_k(\theta_k, \hat{\theta}_k(\vec{Z}))  [\theta_k \mid \vec{Z}] \bigg(\int_{\Theta_{\setminus k}} [\vec{\theta}_{\setminus k} \mid \theta_k, \vec{Z}] \d\vec{\theta}_{\setminus k}\bigg) \d\theta_k \\
&= \sum_{k=1}^p \int_{\Theta_k} L_k(\theta_k, \hat{\theta}_k(\vec{Z}))  [\theta_k \mid \vec{Z}]\d\theta_k,
\end{align*}
which is minimised by minimising $\int_{\Theta_k} L_k(\theta_k, \hat{\theta}_k(\vec{Z}))  [\theta_k \mid \vec{Z}]\d\theta_k$ for each \mbox{$k = 1, \dots, p$}. 
 Hence, for $k = 1, \dots,p$, $\hat{\theta}_k^\star(\cdot)$ is a Bayes estimator with respect to the loss $L_k(\cdot,\cdot)$ for any $\vec{Z} \in \samplespace$. 
\end{proof*}

\bigskip\noindent The estimator $\hat{\theta}_k^\star(\cdot)$, $k = 1, \dots,p$, is hence a functional of the marginal posterior distribution of $\theta_k$, where the functional is the usual Bayes estimator with respect to $L_k(\cdot, \cdot)$. For example, if $L_k(\cdot, \cdot)$ is the absolute-error loss, then $\hat{\theta}_k^\star(\cdot)$ is the marginal posterior median of $\theta_k$.

\bibliographystyle{apalike} 
{\small
\bibliography{bibliography}}

\begin{thebibliography}{}

\bibitem[Cooley et~al., 2006]{Cooley2006}
Cooley, D., Naveau, P., and Poncet, P. (2006).
\newblock Variograms for spatial max-stable random fields.
\newblock In Bertail, P., Soulier, P., and Doukhan, P., editors, {\em
  Dependence in Probability and Statistics}, pages 373--390. Springer, New
  York, NY.

\bibitem[Davison et~al., 2012]{Davison_2012_spatial_extremes}
Davison, A.~C., Padoan, S.~A., and Ribatet, M. (2012).
\newblock Statistical modeling of spatial extremes.
\newblock {\em Statistical Science}, 27:161--186.

\bibitem[Gerber and Nychka, 2021]{Gerber_Nychka_2021_NN_param_estimation}
Gerber, F. and Nychka, D.~W. (2021).
\newblock Fast covariance parameter estimation of spatial {Gaussian} process
  models using neural networks.
\newblock {\em Stat}, 10:e382.

\bibitem[Guinness, 2018]{Guinness_2018}
Guinness, J. (2018).
\newblock Permutation and grouping methods for sharpening {G}aussian process
  approximations.
\newblock {\em Technometrics}, 60:415--429.

\bibitem[Illian et~al., 2008]{Illian_2008_point_patterns}
Illian, J., Penttinen, A., Stoyan, H., and Stoyan, D. (2008).
\newblock {\em Statistical Analysis and Modelling of Spatial Point Patterns}.
\newblock Wiley, New York, NY.

\bibitem[Matheron, 1987]{Matheron_1987}
Matheron, G. (1987).
\newblock Suffit-il, pour une covariance, d’\^{e}tre de type positif.
\newblock {\em Sciences de la Terre, S\'{e}rie Informatique G\'{e}ologique},
  26:51--66.

\bibitem[Naveau et~al., 2009]{Naveau_2009}
Naveau, P., Guillou, A., Cooley, D., and Diebolt, J. (2009).
\newblock Modelling pairwise dependence of maxima in space.
\newblock {\em Biometrika}, 96:1--17.

\end{thebibliography}


\begin{thebibliography}{}

\bibitem[Alcántara et~al., 2023]{Alcantara_2023}
Alcántara, A., Galván, I., and Aler, R. (2023).
\newblock Deep neural networks for the quantile estimation of regional
  renewable energy production.
\newblock {\em Applied Intelligence}, 53:8318--8353.

\bibitem[Baddeley et~al., 2015]{Baddeley_2015_point_patterns}
Baddeley, A., Rubak, E., and Turner, R. (2015).
\newblock {\em Spatial Point Patterns: Methodology and Applications with {R}}.
\newblock Chapman \& Hall/CRC, Boca Raton, FL.

\bibitem[Banesh et~al., 2021]{Banesh_2021_neural_estimator_GP}
Banesh, D., Panda, N., Biswas, A., Roekel, L.~V., Oyen, D., Urban, N.,
  Grosskopf, M., Wolfe, J., and Lawrence, E. (2021).
\newblock Fast {G}aussian process estimation for large-scale in situ inference
  using convolutional neural networks.
\newblock In Chen, Y., Ludwig, H., Tu, Y., Fayyad, U., Zhu, X., Hu, X., Byna,
  S., Liu, X., Zhang, J., Pan, S., Papalexakis, V., Wang, J., Cuzzocrea, A.,
  and Ordonez, C., editors, {\em {IEEE} International Conference on Big Data
  (2021)}, pages 3731--3739. IEEE.
\newblock \url{https://doi.org/10.1109/BigData52589.2021.9671929}.

\bibitem[Bassett and Koenker, 1982]{Bassett_Koener_1982}
Bassett, G. and Koenker, R. (1982).
\newblock An empirical quantile function for linear models with {iid} errors.
\newblock {\em Journal of the American Statistical Association}, 77:407--415.

\bibitem[Battaglia et~al., 2018]{Battaglia_2018}
Battaglia, P.~W., Hamrick, J.~B., Bapst, V., Sanchez{-}Gonzalez, A., Zambaldi,
  V.~F., Malinowski, M., et~al. (2018).
\newblock Relational inductive biases, deep learning, and graph networks.
\newblock {\em arXiv:1806.01261}.

\bibitem[Bevilacqua et~al., 2012]{Bevilacqua_2012}
Bevilacqua, M., Gaetan, C., Mateu, J., and Porcu, E. (2012).
\newblock Estimating space and space-time covariance functions for large data
  sets: A weighted composite likelihood approach.
\newblock {\em Journal of the American Statistical Association}, 107:268--280.

\bibitem[Bronstein et~al., 2017]{Bronstein_2017}
Bronstein, M.~M., Bruna, J., LeCun, Y., Szlam, A., and Vandergheynst, P.
  (2017).
\newblock Geometric deep learning: Going beyond {E}uclidean data.
\newblock {\em IEEE Signal Processing Magazine}, 34:18--42.

\bibitem[Brown and Purves, 1973]{Brown_1973}
Brown, L.~D. and Purves, R. (1973).
\newblock Measurable selections of extrema.
\newblock {\em The Annals of Statistics}, 1:902--912.

\bibitem[Cannon, 2011]{Cannon_2011_quantile_regression_neural_networks}
Cannon, A.~J. (2011).
\newblock Quantile regression neural networks: Implementation in {R} and
  application to precipitation downscaling.
\newblock {\em Computers \& Geosciences}, 37:1277--1284.

\bibitem[Cannon, 2018]{Cannon_2018}
Cannon, A.~J. (2018).
\newblock Non-crossing nonlinear regression quantiles by monotone composite
  quantile regression neural network.
\newblock {\em Stochastic Environmental Research and Risk Assessment},
  32:3207--3225.

\bibitem[Canty et~al., 2006]{Canty_2006}
Canty, A.~J., Davison, A.~C., Hinkley, D.~V., and Ventura, V. (2006).
\newblock Bootstrap diagnostics and remedies.
\newblock {\em The Canadian Journal of Statistics}, 34:5--27.

\bibitem[Cao et~al., 2013]{Cao_2013_Suomi_NPP}
Cao, C., Xiong, J., Blonski, S., Liu, Q., Uprety, S., Shao, X., Bai, Y., and
  Weng, F. (2013).
\newblock Suomi {NPP} {VIIRS} sensor data record verification, validation, and
  long-term performance monitoring.
\newblock {\em Journal of Geophysical Research: Atmospheres}, 118:11--664.

\bibitem[Castro-Camilo and Huser, 2020]{Castro-Camilo_2020_local_estimation}
Castro-Camilo, D. and Huser, R. (2020).
\newblock Local likelihood estimation of complex tail dependence structures,
  applied to {U.S.} precipitation extremes.
\newblock {\em Journal of the American Statistical Association},
  115:1037--1054.

\bibitem[Castruccio et~al.,
  2016]{Castruccio_2016_composite_likelihood_max-stable}
Castruccio, S., Huser, R., and Genton, M.~G. (2016).
\newblock High-order composite likelihood inference for max-stable
  distributions and processes.
\newblock {\em Journal of Computational and Graphical Statistics},
  25:1212--1229.

\bibitem[Chan et~al., 2018]{Chan_2018}
Chan, J., Perrone, V., Spence, J., Jenkins, P., Mathieson, S., and Song, Y.
  (2018).
\newblock A likelihood-free inference framework for population genetic data
  using exchangeable neural networks.
\newblock In Bengio, S., Wallach, H., Larochelle, H., Grauman, K.,
  Cesa-Bianchi, N., and Garnett, R., editors, {\em Advances in Neural
  Information Processing Systems}, volume~31. Curran Associates, Inc., Red
  Hook, NY.

\bibitem[Chen et~al.,
  2021]{Chen_2021_NN-approximate_sufficient_statistics_for_implicit_models}
Chen, Y., Zhang, D., Gutmann, M.~U., Courville, A., and Zhu, Z. (2021).
\newblock Neural approximate sufficient statistics for implicit models.
\newblock In Qian, Y., Tan, Z., Sun, X., Lin, M., Li, D., Sun, Z., Li, H., and
  Jin, R., editors, {\em Proceedings of the 9th International Conference on
  Learning Representations (ICLR 2021)}.
\newblock Virtual: OpenReview. \url{https://openreview.net/pdf?id=SRDuJssQud}.

\bibitem[Chernozhukov et~al., 2010]{Chernozhukov_2010}
Chernozhukov, V., Fernández-Val, I., and Galichon, A. (2010).
\newblock Quantile and probability curves without crossing.
\newblock {\em Econometrica}, 78:1093--1125.

\bibitem[Cisneros et~al., 2024]{Cisneros_2023_GNNs_Australian_wildfire}
Cisneros, D., Richards, J., Dahal, A., Lombardo, L., and Huser, R. (2024).
\newblock Deep graphical regression for jointly moderate and extreme
  {A}ustralian wildfires.
\newblock {\em Spatial Statistics}, 59:100811.

\bibitem[Creel, 2017]{Creel_2017}
Creel, M. (2017).
\newblock Neural nets for indirect inference.
\newblock {\em Econometrics and Statistics}, 2:36--49.

\bibitem[Cressie et~al.,
  2021]{Cressie_2021_review_spatial-basis-function_models}
Cressie, N., Sainsbury-Dale, M., and Zammit-Mangion, A. (2021).
\newblock Basis-function models in spatial statistics.
\newblock {\em Annual Review of Statistics and its Applications}, 9:373--400.

\bibitem[Danel et~al., 2020]{Danel_2020_spatial_GNN}
Danel, T., Spurek, P., Tabor, J., {\'{S}}mieja, M., Struski, {\L}., S{\l}owik,
  A., and Maziarka, {\L}. (2020).
\newblock Spatial graph convolutional networks.
\newblock In Yang, H., Pasupa, K., Leung, A. C.-S., Kwok, J.~T., Chan, J.~H.,
  and King, I., editors, {\em Proceedings of the 27th International Conference
  on Neural Information Processing (ICONIP 2020)}, pages 668--675. Springer,
  Cham.

\bibitem[Davison and Huser, 2015]{Davison_Huser_2015_Statistics_of_extremes}
Davison, A.~C. and Huser, R. (2015).
\newblock Statistics of extremes.
\newblock {\em Annual Review of Statistics and its Application}, 2:203--235.

\bibitem[Davison et~al., 2019]{Davison_2019_spatial_extremes}
Davison, A.~C., Huser, R., and Thibaud, E. (2019).
\newblock Spatial extremes.
\newblock In Gelfand, A.~E., Fuentes, M., Hoeting, J.~A., and Smith, R.~L.,
  editors, {\em Handbook of Environmental and Ecological Statistics}, pages
  711--744. Chapman \& Hall/CRC Press, Boca Raton, FL.

\bibitem[Diggle, 2013]{Diggle_2013_point_patterns}
Diggle, P.~J. (2013).
\newblock {\em Statistical Analysis of Spatial and Spatio-Temporal Point
  Patterns}.
\newblock Chapman \& Hall/CRC, New York, NY, 3rd edition.

\bibitem[Dyer et~al., 2022]{Dyer_2022_GNN}
Dyer, J., Cannon, P., Doyne~Farmer, J., and Schmon, S.~M. (2022).
\newblock Calibrating agent-based models to microdata with graph neural
  networks.
\newblock {\em arXiv:2206.07570}.

\bibitem[Fisher et~al., 2023]{Fisher_2023_neural_quantile_regression}
Fisher, T., Luedtke, A., Carone, M., and Simon, N. (2023).
\newblock Marginal {B}ayesian posterior inference using recurrent neural
  networks with application to sequential models.
\newblock {\em Statistica Sinica}, 33:1507--1532.

\bibitem[Flagel et~al., 2018]{Flagel_2018}
Flagel, L., Brandvain, Y., and Schrider, D.~R. (2018).
\newblock The unreasonable effectiveness of convolutional neural networks in
  population genetic inference.
\newblock {\em Molecular Biology and Evolution}, 36:220--238.

\bibitem[Genton and Kleiber, 2015]{Genton_2015_multivariate_geostatistics}
Genton, M.~G. and Kleiber, W. (2015).
\newblock Cross-covariance functions for multivariate geostatistics.
\newblock {\em Statistical Science}, 30:147--163.

\bibitem[Genton et~al., 2015]{Genton_2015_multivariate_maxstable_processes}
Genton, M.~G., Padoan, S.~A., and Sang, H. (2015).
\newblock Multivariate max-stable spatial processes.
\newblock {\em Biometrika}, 102:215--230.

\bibitem[Gerber and Nychka, 2021]{Gerber_Nychka_2021_NN_param_estimation}
Gerber, F. and Nychka, D.~W. (2021).
\newblock Fast covariance parameter estimation of spatial {Gaussian} process
  models using neural networks.
\newblock {\em Stat}, 10:e382.

\bibitem[Gilmer et~al., 2017]{Gilmer_2017_message_passing_GNNs}
Gilmer, J., Schoenholz, S.~S., Riley, P.~F., Vinyals, O., and Dahl, G.~E.
  (2017).
\newblock Neural message passing for quantum chemistry.
\newblock In Precup, D. and Teh, Y.~W., editors, {\em Proceedings of the 34th
  International Conference on Machine Learning (ICML 2017)}, pages 1263--1272.
  PMLR.

\bibitem[Gl{\"o}ckler et~al., 2022]{Glockler_2022}
Gl{\"o}ckler, M., Deistler, M., and Macke, J.~H. (2022).
\newblock Variational methods for simulation-based inference.
\newblock In {\em Proceedings of the 10th International Conference on Learning
  Representations (ICLR 2022)}.
\newblock Virtual: OpenReview.
  \url{https://openreview.net/forum?id=kZ0UYdhqkNY}.

\bibitem[Gneiting et~al., 2010]{Gneiting_2010_multivariate_Matern}
Gneiting, T., Kleiber, W., and Schlather, M. (2010).
\newblock {Matérn} cross-covariance functions for multivariate random fields.
\newblock {\em Journal of the American Statistical Association},
  105:1167--1177.

\bibitem[Gonçalves et~al., 2020]{Goncalves_2020}
Gonçalves, P.~J., Lueckmann, J.-M., Deistler, M., Nonnenmacher, M., Öcal, K.,
  Bassetto, G., Chintaluri, C., Podlaski, W.~F., Haddad, S.~A., Vogels, T.~P.,
  Greenberg, D.~S., and Macke, J.~H. (2020).
\newblock Training deep neural density estimators to identify mechanistic
  models of neural dynamics.
\newblock {\em eLife}, 9:e56261.

\bibitem[Goodfellow et~al., 2016]{Goodfellow_2016_Deep_learning}
Goodfellow, I., Bengio, Y., and Courville, A. (2016).
\newblock {\em {Deep Learning}}.
\newblock MIT Press, Cambridge, MA.

\bibitem[Grattarola et~al., 2022]{Grattarola_2022_pooling}
Grattarola, D., Zambon, D., Bianchi, F.~M., and Alippi, C. (2022).
\newblock Understanding pooling in graph neural networks.
\newblock {\em IEEE Transactions on Neural Networks and Learning Systems},
  35:2708--2718.

\bibitem[Greenberg et~al., 2019]{Greenberg_2019}
Greenberg, D., Nonnenmacher, M., and Macke, J. (2019).
\newblock Automatic posterior transformation for likelihood-free inference.
\newblock In Chaudhuri, K. and Salakhutdinov, R., editors, {\em Proceedings of
  the 36th International Conference on Machine Learning (ICML 2019)}, pages
  2404--2414. PMLR.

\bibitem[Gupta et~al., 2016]{Gupta_2016_monotonic_networks}
Gupta, M., Cotter, A., Pfeifer, J., Voevodski, K., Canini, K., Mangylov, A.,
  Moczydlowski, W., and van Esbroeck, A. (2016).
\newblock Monotonic calibrated interpolated look-up tables.
\newblock {\em Journal of Machine Learning Research}, 17:1--47.

\bibitem[Haas, 1990a]{Haas_1990_moving_window}
Haas, T.~C. (1990a).
\newblock Kriging and automated variogram modeling within a moving window.
\newblock {\em Atmospheric Environment}, 24:1759--1769.

\bibitem[Haas, 1990b]{Haas_1990_moving_window_lognormal}
Haas, T.~C. (1990b).
\newblock Lognormal and moving window methods of estimating acid deposition.
\newblock {\em Journal of the American Statistical Association}, 85:950--963.

\bibitem[Han et~al.,
  2022]{Han_2019_universal_approximation_of_symmetric_functions}
Han, J., Li, Y., Lin, L., Lu, J., Zhang, J., and Zhang, L. (2022).
\newblock Universal approximation of symmetric and anti-symmetric functions.
\newblock {\em Communications in Mathematical Sciences}, 20:1397--1408.

\bibitem[He et~al., 2014]{He_2014_spatial_pyramid_pooling}
He, K., Zhang, X., Ren, S., and Sun, J. (2014).
\newblock Spatial pyramid pooling in deep convolutional networks for visual
  recognition.
\newblock In Fleet, D., Pajdla, T., Schiele, B., and Tuytelaars, T., editors,
  {\em Computer Vision (ECCV 2014)}, pages 346--361. Springer, Cham.

\bibitem[He, 1997]{He_1997_quantile_curves_without_crossing}
He, X. (1997).
\newblock Quantile curves without crossing.
\newblock {\em The American Statistician}, 51:186--192.

\bibitem[Hermans et~al., 2020]{Hermans_2020}
Hermans, J., Begy, V., and Louppe, G. (2020).
\newblock Likelihood-free {MCMC} with amortized approximate ratio estimators.
\newblock In III, H.~D. and Singh, A., editors, {\em Proceedings of the 37th
  International Conference on Machine Learning}, pages 4239--4248. PMLR.

\bibitem[Hermans et~al., 2022]{Hermans_2022}
Hermans, J., Delaunoy, A., Rozet, F., Wehenkel, A., Begy, V., and Louppe, G.
  (2022).
\newblock A crisis in simulation-based inference? {B}eware, your posterior
  approximations can be unfaithful.
\newblock {\em Transactions on Machine Learning Research.}
\newblock OpenReview. \url{https://openreview.net/pdf?id=LHAbHkt6Aq}.

\bibitem[Hjelm et~al., 2016]{Hjelm_2016}
Hjelm, D., Salakhutdinov, R.~R., Cho, K., Jojic, N., Calhoun, V., and Chung, J.
  (2016).
\newblock Iterative refinement of the approximate posterior for directed belief
  networks.
\newblock In {\em Proceedings of the 30th Conference on Neural Information
  Processing Systems}, pages 4698--4706, Red Hook, NY. Curran.

\bibitem[Hornik et~al., 1989]{Hornik_1989_FNN_universal_approximation_theorem}
Hornik, K., Stinchcombe, M., and White, H. (1989).
\newblock Multilayer feedforward networks are universal approximators.
\newblock {\em Neural Networks}, 2:359--366.

\bibitem[Huser, 2013]{Huser_2013_PhD_thesis}
Huser, R. (2013).
\newblock {\em Statistical Modeling and Inference for Spatio-Temporal
  Extremes}.
\newblock PhD thesis, Swiss Federal Institute of Technology, Lausanne,
  Switzerland.

\bibitem[Huser et~al., 2019]{Huser_2019_advances_in_spatial_extremes}
Huser, R., Dombry, C., Ribatet, M., and Genton, M.~G. (2019).
\newblock Full likelihood inference for max-stable data.
\newblock {\em Stat}, 8:e218.

\bibitem[Huser et~al., 2024]{Huser_2024_max-stable}
Huser, R., Opitz, T., and Wadsworth, J. (2024).
\newblock Modeling of spatial extremes in environmental data science: Time to
  move away from max-stable processes.
\newblock {\em arXiv:2401.17430}.

\bibitem[Huser and Wadsworth, 2022]{Huser_2022_advances_in_spatial_extremes}
Huser, R. and Wadsworth, J. (2022).
\newblock Advances in statistical modeling of spatial extremes.
\newblock {\em Wiley Interdisciplinary Reviews: Computational Statistics},
  14:e1537.

\bibitem[Illian et~al., 2008]{Illian_2008_point_patterns}
Illian, J., Penttinen, A., Stoyan, H., and Stoyan, D. (2008).
\newblock {\em Statistical Analysis and Modelling of Spatial Point Patterns}.
\newblock Wiley, New York, NY.

\bibitem[Jiang et~al., 2017]{Jiang_2017}
Jiang, B., Wu, T.-Y., Zheng, C., and Wong, W.~H. (2017).
\newblock Learning summary statistic for approximate {B}ayesian computation via
  deep neural network.
\newblock {\em Statistica Sinica}, 27:1595--1618.

\bibitem[Klemmer et~al., 2023]{Klemmer_2023_spatial_GNN}
Klemmer, K., Safir, N.~S., and Neill, D.~B. (2023).
\newblock Positional encoder graph neural networks for geographic data.
\newblock In Ruiz, F., Dy, J., and van~de Meent, J.-W., editors, {\em
  Proceedings of The 26th International Conference on Artificial Intelligence
  and Statistics}, pages 1379--1389. PMLR.

\bibitem[Koenker and Bassett, 1978]{Koenker_1978_quantile_regression}
Koenker, R. and Bassett, G. (1978).
\newblock Regression quantiles.
\newblock {\em Econometrica}, 46:33--50.

\bibitem[Koenker and Hallock, 2001]{Koenker_2001_quantile_regression}
Koenker, R. and Hallock, K.~F. (2001).
\newblock Quantile regression.
\newblock {\em Journal of Economic Perspectives}, 15:143--156.

\bibitem[Kuusela and Stein, 2018]{Kuusela_Stein_2018}
Kuusela, M. and Stein, M.~L. (2018).
\newblock Locally stationary spatio-temporal interpolation of {A}rgo profiling
  float data.
\newblock {\em Proceedings of the Royal Society A}, 474:1--24.

\bibitem[Lehmann and Casella, 1998]{Lehmann_Casella_1998_Point_Estimation}
Lehmann, E.~L. and Casella, G. (1998).
\newblock {\em Theory of Point Estimation}.
\newblock Springer, New York, NY, 2nd edition.

\bibitem[Lenzi et~al., 2023]{Lenzi_2021_NN_param_estimation}
Lenzi, A., Bessac, J., Rudi, J., and Stein, M.~L. (2023).
\newblock Neural networks for parameter estimation in intractable models.
\newblock {\em Computational Statistics \& Data Analysis}, 185:107762.

\bibitem[Lucibello, 2021]{Lucibello2021GNN}
Lucibello, C. (2021).
\newblock {G}raph{N}eural{N}etworks.jl: a geometric deep learning library for
  the {J}ulia programming language.
\newblock \url{https://github.com/CarloLucibello/GraphNeuralNetworks.jl}.

\bibitem[Lusher et~al., 2013]{Lusher_2013_ERGMs}
Lusher, D., Koskinen, J., and Robins, G. (2013).
\newblock {\em Exponential Random Graph Models for Social Networks: Theory,
  Methods, and Applications}.
\newblock Cambridge University Press, Cambridge, UK.

\bibitem[Madrid-Padilla et~al., 2022]{Madrid-Padilla_2022}
Madrid-Padilla, O.~H., Tansey, W., and Chen, Y. (2022).
\newblock Quantile regression with {ReLU} networks: Estimators and minimax
  rates.
\newblock {\em Journal of Machine Learning Research}, 23:1--42.

\bibitem[Mesquita et~al., 2020]{Mesquita_2020_pooling}
Mesquita, D., Souza, A.~H., and Kaski, S. (2020).
\newblock Rethinking pooling in graph neural networks.
\newblock In {\em Proceedings of the 34th International Conference on Neural
  Information Processing Systems}, pages 2220--2231. Curran Associates Inc.,
  Red Hook, NY.

\bibitem[Mestdagh et~al., 2019]{Mestdagh_2019_prepaid_ABC}
Mestdagh, M., Verdonck, S., Meers, K., Loossens, T., and Tuerlinckx, F. (2019).
\newblock Prepaid parameter estimation without likelihoods.
\newblock {\em PLoS Computational Biology}, 15:e1007181.

\bibitem[Møller and Waagepetersen, 2004]{Moller_2004_point_patterns}
Møller, J. and Waagepetersen, R.~P. (2004).
\newblock {\em Statistical Inference and Simulation for Spatial Point
  Processes}.
\newblock Chapman \& Hall/CRC, Boca Raton, FL.

\bibitem[Navarin et~al., 2019]{Navarin_2019_universal_readout}
Navarin, N., Tran, D.~V., and Sperduti, A. (2019).
\newblock Universal readout for graph convolutional neural networks.
\newblock In {\em 2019 International Joint Conference on Neural Networks
  (IJCNN)}, pages 1--7. IEEE.

\bibitem[Pacchiardi and Dutta, 2022]{Pacchiardi_2022_GANs_scoring_rules}
Pacchiardi, L. and Dutta, R. (2022).
\newblock Likelihood-free inference with generative neural networks via scoring
  rule minimization.
\newblock {\em arXiv:2205.15784}.

\bibitem[Padoan et~al.,
  2010]{Padoan_2010_composite_likelihood_max_stable_processes}
Padoan, S.~A., Ribatet, M., and Sisson, S.~A. (2010).
\newblock Likelihood-based inference for max-stable processes.
\newblock {\em Journal of the American Statistical Association}, 105:263--277.

\bibitem[Papamakarios et~al., 2019]{Papamakarios_2019}
Papamakarios, G., Sterratt, D., and Murray, I. (2019).
\newblock Sequential neural likelihood: Fast likelihood-free inference with
  autoregressive flows.
\newblock In Chaudhuri, K. and Sugiyama, M., editors, {\em Proceedings of the
  Twenty-Second International Conference on Artificial Intelligence and
  Statistics}, pages 837--848. PMLR.

\bibitem[Pasche and Engelke, 2024]{Pasche_Engelke_2022}
Pasche, O.~C. and Engelke, S. (2024).
\newblock Neural networks for extreme quantile regression with an application
  to forecasting of flood risk.
\newblock {\em Annals of Applied Statistics}, 18:2818--2839.

\bibitem[Pfreundschuh et~al., 2018]{Pfreundschuh_2018}
Pfreundschuh, S., Eriksson, P., Duncan, D., Rydberg, B., H{\aa}kansson, N., and
  Thoss, A. (2018).
\newblock A neural network approach to estimating a posteriori distributions of
  {B}ayesian retrieval problems.
\newblock {\em Atmospheric Measurement Techniques}, 11:4627--4643.

\bibitem[Radev et~al., 2022]{Radev_2022_BayesFlow}
Radev, S.~T., Mertens, U.~K., Voss, A., Ardizzone, L., and K\"othe, U. (2022).
\newblock {BayesFlow}: Learning complex stochastic models with invertible
  neural networks.
\newblock {\em IEEE Transactions on Neural Networks and Learning Systems},
  33:1452--1466.

\bibitem[Radev et~al., 2023]{Radev_2023_JANA}
Radev, S.~T., Schmitt, M., Pratz, V., Picchini, U., K\"othe, U., and B\"urkner,
  P.-C. (2023).
\newblock {JANA}: Jointly amortized neural approximation of complex {B}ayesian
  models.
\newblock In Evans, R.~J. and Shpitser, I., editors, {\em Proceedings of the
  Thirty-Ninth Conference on Uncertainty in Artificial Intelligence}, pages
  1695--1706. PMLR.

\bibitem[Rai et~al., 2024]{Rai_2023}
Rai, S., Hoffman, A., Lahiri, S., Nychka, D.~W., Sain, S.~R., and
  Bandyopadhyay, S. (2024).
\newblock Fast parameter estimation of generalized extreme value distribution
  using neural networks.
\newblock {\em Environmetrics}, 35:e2845.

\bibitem[Richards and Huser,
  2022]{Richards_Huser_2022_PINN_quantile_regression}
Richards, J. and Huser, R. (2022).
\newblock Regression modelling of spatiotemporal extreme {US} wildfires via
  partially-interpretable neural networks.
\newblock {\em arXiv:2208.07581}.

\bibitem[Richards et~al., 2025]{Richards_2023_censoring}
Richards, J., Sainsbury-Dale, M., Huser, R., and Zammit-Mangion, A. (2025).
\newblock Neural {B}ayes estimators for censored inference with
  peaks-over-threshold models.
\newblock {\em Journal of Machine Learning Research}, to appear.
\newblock \url{https://doi.org/10.48550/arXiv.2306.15642}.

\bibitem[Robert, 2007]{Robert_2007_The_Bayesian_Choice}
Robert, C.~P. (2007).
\newblock {\em The {B}ayesian Choice}.
\newblock Springer, New York, NY, 2nd edition.

\bibitem[Robins et~al., 2007]{Robins_2007_ERGMs}
Robins, G., Pattison, P., Kalish, Y., and Lusher, D. (2007).
\newblock An introduction to exponential random graph $(p^*)$ models for social
  networks.
\newblock {\em Social Networks}, 29:173--191.

\bibitem[Rudi et~al., 2021]{Rudi_2020_NN_parameter_estimation}
Rudi, J., Julie, B., and Lenzi, A. (2021).
\newblock Parameter estimation with dense and convolutional neural networks
  applied to the {FitzHugh-Nagumo ODE}.
\newblock In Bruna, J., Hesthaven, J., and Zdeborova, L., editors, {\em
  Proceedings of the 2nd Annual Conference on Mathematical and Scientific
  Machine Learning}, pages 1--28. PMLR.

\bibitem[Sainsbury-Dale, 2024]{NeuralEstimators}
Sainsbury-Dale, M. (2024).
\newblock {\em {NeuralEstimators}: Likelihood-Free Parameter Estimation using
  Neural Networks}.
\newblock R package version 0.1.2,
  \url{https://CRAN.R-project.org/package=NeuralEstimators}.

\bibitem[Sainsbury-Dale et~al., 2025]{Sainsbury_2025}
Sainsbury-Dale, M., Zammit-Mangion, A., Cressie, N., and Huser, R. (2025).
\newblock Neural parameter estimation with incomplete data.
\newblock {\em arXiv:2501.04330}.

\bibitem[Sainsbury-Dale et~al.,
  2024]{Sainsbury-Dale_2022_neural_Bayes_estimators}
Sainsbury-Dale, M., Zammit-Mangion, A., and Huser, R. (2024).
\newblock Likelihood-free parameter estimation with neural {B}ayes estimators.
\newblock {\em The American Statistician}, 78:1--14.

\bibitem[Sang and Genton, 2012]{Sang_Genton_2014}
Sang, H. and Genton, M.~G. (2012).
\newblock Tapered composite likelihood for spatial max-stable models.
\newblock {\em Spatial Statistics}, 8:86--103.

\bibitem[Schlather, 2002]{Schlather_2002_max-stable_models}
Schlather, M. (2002).
\newblock Models for stationary max-stable random fields.
\newblock {\em Extremes}, 5:33--44.

\bibitem[Sill, 1997]{Sill_1997_Monotonic_networks}
Sill, J. (1997).
\newblock Monotonic networks.
\newblock In Jordan, M., Kearns, M., and Solla, S., editors, {\em Advances in
  Neural Information Processing Systems}, volume~10, pages 661--667. MIT Press.

\bibitem[Taylor, 2000]{Taylor_2000_quantile_regression}
Taylor, J.~W. (2000).
\newblock A quantile regression neural network approach to estimating the
  conditional density of multiperiod returns.
\newblock {\em Journal of Forecasting}, 19:299--311.

\bibitem[Thomas et~al., 2022]{Thomas_2022_ratio_estimation}
Thomas, O., Dutta, R., Corander, J., Kaski, S., and Gutmann, M.~U. (2022).
\newblock Likelihood-free inference by ratio estimation.
\newblock {\em {B}ayesian Analysis}, 17:1--31.

\bibitem[Tonks et~al., 2024]{Tonks_2022_GNNs_geostats}
Tonks, A., Harris, T., Li, B., Brown, W., and Smith, R. (2024).
\newblock Forecasting {W}est {N}ile virus with graph neural networks:
  Harnessing spatial dependence in irregularly sampled geospatial data.
\newblock {\em GeoHealth}, 8:e2023GH000784.

\bibitem[Tsyrulnikov and Sotskiy, 2023]{Tsyrulnikov_2024}
Tsyrulnikov, M. and Sotskiy, A. (2023).
\newblock Regularization of the ensemble {K}alman filter using a
  non-parametric, non-stationary spatial model.
\newblock {\em arXiv:2306.14318}.

\bibitem[Wagstaff et~al.,
  2022]{Wagstaff_2021_universal_approximation_set_functions}
Wagstaff, E., Fuchs, F.~B., Engelcke, M., Osborne, M., and Posner, I. (2022).
\newblock Universal approximation of functions on sets.
\newblock {\em Journal of Machine Learning Research}, 23:1--56.

\bibitem[Walchessen et~al., 2024]{Walchessen_2023_neural_likelihood_surfaces}
Walchessen, J., Lenzi, A., and Kuusela, M. (2024).
\newblock Neural likelihood surfaces for spatial processes with computationally
  intensive or intractable likelihoods.
\newblock {\em Spatial Statistics}, 62:100848.

\bibitem[Wehenkel and Louppe, 2019]{Wehenkel_Louppe_2019_monotonic_networks}
Wehenkel, A. and Louppe, G. (2019).
\newblock Unconstrained monotonic neural networks.
\newblock In Wallach, H., Larochelle, H., Beygelzimer, A., d\textquotesingle
  Alch\'{e}-Buc, F., Fox, E., and Garnett, R., editors, {\em 33rd Conference on
  Neural Information Processing Systems}, pages 1543--1553, Red Hook, NY.
  Curran.

\bibitem[Winkler et~al., 2019]{Winkler_2019_likelihood-free_normalising_flows}
Winkler, C., Worrall, D.~E., Hoogeboom, E., and Welling, M. (2019).
\newblock Learning likelihoods with conditional normalizing flows.
\newblock {\em arXiv:1912.00042}.

\bibitem[Wiqvist et~al., 2021]{Wiqvist_2021}
Wiqvist, S., Frellsen, J., and Picchini, U. (2021).
\newblock Sequential neural posterior and likelihood approximation.
\newblock {\em arXiv:2102.06522}.

\bibitem[Wu et~al., 2021]{Wu_2021_GNN_review}
Wu, Z., Pan, S., Chen, F., Long, G., Zhang, C., and Yu, P.~S. (2021).
\newblock A comprehensive survey on graph neural networks.
\newblock {\em IEEE Transactions on Neural Networks and Learning Systems},
  32:4--24.

\bibitem[Xu and Raginsky, 2022]{Xu_Raginsky_2022}
Xu, A. and Raginsky, M. (2022).
\newblock Minimum excess risk in {B}ayesian learning.
\newblock {\em IEEE Transactions on Information Theory}, 68:7935--7955.

\bibitem[Xu et~al., 2017]{Xu_2017_quantile_regression_neural_network}
Xu, Q., Deng, K., Jiang, C., Sun, F., and Huang, X. (2017).
\newblock Composite quantile regression neural network with applications.
\newblock {\em Expert Systems with Applications}, 76:129--139.

\bibitem[Yu and Moyeed, 2001]{Yu_Moyeed_2001}
Yu, K. and Moyeed, R.~A. (2001).
\newblock {B}ayesian quantile regression.
\newblock {\em Statistics \& Probability Letters}, 54:437--447.

\bibitem[Zaheer et~al., 2017]{Zaheer_2017_Deep_Sets}
Zaheer, M., Kottur, S., Ravanbakhsh, S., Poczos, B., Salakhutdinov, R.~R., and
  Smola, A.~J. (2017).
\newblock Deep sets.
\newblock In {\em Proceedings of the 31st Conference on Neural Information
  Processing Systems}, pages 3392--3402, Red Hook, NY. Curran.

\bibitem[Zammit-Mangion and Rougier, 2020]{Zammit_2020}
Zammit-Mangion, A. and Rougier, J. (2020).
\newblock Multi-scale process modelling and distributed computation for spatial
  data.
\newblock {\em Statistics and Computing}, 30:1609--1627.

\bibitem[Zammit-Mangion et~al., 2025]{Zammit_2024_ARSIA}
Zammit-Mangion, A., Sainsbury-Dale, M., and Huser, R. (2025).
\newblock Neural methods for amortized inference.
\newblock {\em Annual Review of Statistics and Its Application}, to appear.
\newblock \url{https://doi.org/10.48550/arXiv.2404.12484}.

\bibitem[Zammit-Mangion and Wikle, 2020]{Zammit-Mangion_Wikle_2020}
Zammit-Mangion, A. and Wikle, C.~K. (2020).
\newblock Deep integro-difference equation models for spatio-temporal
  forecasting.
\newblock {\em Spatial Statistics}, 37:100408.

\bibitem[Zhan and Datta, 2024]{Zhan_Datta_2023_GNNs_geostats}
Zhan, W. and Datta, A. (2024).
\newblock Neural networks for geospatial data.
\newblock {\em Journal of the American Statistical Association}, to appear.
\newblock \url{https://doi.org/10.48550/arXiv:2304.09157}.

\bibitem[Zhang et~al., 2018]{Zhang_2018_SortPool}
Zhang, M., Cui, Z., Neumann, M., and Chen, Y. (2018).
\newblock An end-to-end deep learning architecture for graph classification.
\newblock In {\em Proceedings of the 32nd AAAI Conference on Artificial
  Intelligence (AAAI 2018)}, pages 4438--4445. AAAI Press.

\bibitem[Zhang et~al., 2019]{Zhang_2019_GCN_review}
Zhang, S., Tong, H., Xu, J., and Maciejewski, R. (2019).
\newblock Graph convolutional networks: A comprehensive review.
\newblock {\em Computational Social Networks}, 6:1--23.

\bibitem[Zhang and Zhao, 2021]{Zhang_2021_spatial_GNN}
Zhang, Z. and Zhao, L. (2021).
\newblock Representation learning on spatial networks.
\newblock In Ranzato, M., Beygelzimer, A., Dauphin, Y., Liang, P., and Vaughan,
  J.~W., editors, {\em Advances in Neural Information Processing Systems},
  volume~34, pages 2303--2318. Curran Associates, Inc., Red Hook, NY.

\bibitem[Zhong and Wang,
  2023]{Zhong_Wang_2023_partially_interpretable_neural_quantile_regression}
Zhong, Q. and Wang, J.-L. (2023).
\newblock Neural networks for partially linear quantile regression.
\newblock {\em Journal of Business \& Economic Statistics}, 42:603--614.

\bibitem[Zhou, 2018]{Zhou_2018_universal_approximation_CNNs}
Zhou, D. (2018).
\newblock Universality of deep convolutional neural networks.
\newblock {\em Applied and Computational Harmonic Analysis}, 48:787--794.

\bibitem[Zhou et~al., 2020]{Zhou_2020_GNN_review}
Zhou, J., Cui, G., Hu, S., Zhang, Z., Yang, C., Liu, Z., Wang, L., Li, C., and
  Sun, M. (2020).
\newblock Graph neural networks: A review of methods and applications.
\newblock {\em AI Open}, 1:57--81.

\end{thebibliography}

\ifbool{arxiv}{

\begin{singlespace}
\ifbool{arxiv}{%
	\begin{center}
		\supptitle
\end{center}}{\maketitle}
\end{singlespace}

\supplement
\begin{bibunit}[apalike] 

In Section~\ref{sec:variogram}, we use the empirical variogram to motivate the GNN architecture we propose for extracting summary statistics from spatial data. In Section~\ref{sec:neighbourhoods}, we investigate several definitions for the neighbourhood of a node. In Section~\ref{sec:experiment:variablesamplesizes}, we illustrate several properties of our estimator with respect to the distribution $\Omega(\cdot)$ for the spatial locations $S$. In Section~\ref{sec:additionalfigures}, we provide additional figures and tables. 

\section{Spatial summary statistics and the variogram}\label{sec:variogram}

Fundamental to neural parameter inference for general spatial models is the learning of summary statistics from spatial data. Recall from the main text that \cite{Gerber_Nychka_2021_NN_param_estimation} use the empirical variogram as an expert hand-crafted summary statistic, which is then mapped to the parameter space using a multilayer perceptron (MLP). The empirical variogram is ideal for use in isotropic Gaussian process models, since for these models the variogram is a sufficient statistic for the covariance-function parameters. It also serves as a useful starting point from which one may glean important properties for constructing more generally-applicable summary statistics for spatial data. 

Given data $\vec{Z} \equiv (Z_1, \dots, Z_n)'$ observed at locations $S \equiv \{\vec{s}_1, \dots, \vec{s}_n\} \subset \mathcal{D}$, where $\mathcal{D}$ is the spatial domain of interest, the empirical semivariogram at spatial distance $h$ is given by
\begin{equation}\label{eqn:variogram}
    \hat{\gamma}_h = \frac{1}{2|B_h|} \sum_{(i, j) \in B_h} (Z_i - Z_j)^2,
\end{equation}
where $B_h \equiv \left\{(i,j) : \|\boldsymbol{s}_i - \boldsymbol{s}_j\| = h \right\}$ denotes the set of indices of pairs of locations separated by a distance $h$ and $|\cdot|$ denotes set cardinality. (In practice, one typically considers distance ``bins'', but we do not make this explicit for notational convenience.) Now, \eqref{eqn:variogram} is a function of a specific subset of the data, namely, those pairs of observations separated by a distance $h$. However, it may be rewritten as a function of all the available data, namely, 
\begin{equation}\label{eqn:variogram2}
    \hat{\gamma}_h = \frac{1}{2}  \sum_{(i, j) \in A} \frac{w(\vec{s}_i, \vec{s}_j)}{\sum_{(i',j') \in A} w(\vec{s}_{i'}, \vec{s}_{j'})} (Z_i - Z_j)^2,
\end{equation}
where $A \equiv \{(i,j) : i,j = 1, \dots, n\}$ denotes the set of all pairs of indices, $w(\vec{s}_i, \vec{s}_j) = \mathbb{I}((i, j) \in B_h)$ is a spatial weight, and $\mathbb{I}(\cdot)$ denotes the indicator function. This representation shows that the empirical (semi)variogram corresponding to distance $h$ is a spatially-weighted sum over a nonlinear function of the spatial data. Importantly, the spatial weights are (i) normalised to sum to one and (ii) a non-monotonic function of spatial distance. Note that, without normalisation, the value of $\hat{\gamma}_h$ would depend on the specific configuration of the spatial locations $S$ (specifically, on $|B_h|$), and this confounding would make inference difficult in the case that $S$ is allowed to vary between data sets.  

 Motivated by the variogram, we now propose a relatively flexible 
 spatial summary statistic, which serves as a useful building block within a more expressive hierarchical representation (e.g., a GNN). This is given by
\begin{equation}\label{eqn:T}
    T(\vec{Z}, S) = \sum_{(i, j) \in N} \frac{w(\vec{s}_i, \vec{s}_j)}{\sum_{(i',j') \in N} w(\vec{s}_{i'}, \vec{s}_{j'})} \rho(Z_i, Z_j),
\end{equation}
 where $N \subseteq A$, $w(\cdot, \cdot)$ is a user-specified or learnable function (e.g., an MLP) of spatial distance (or spatial lag for anisotropic models), and $\rho(\cdot, \cdot)$ is a learnable function, typically an MLP or a parsimonious parametric function such as $\rho(Z_i, Z_j) = |aZ_i - (1-a)Z_j|^b$ for learnable parameters $a \geq 0$ and $b \geq 0$ \citep[inspired by the so-called ``variogram of order $\alpha$'';][]{Matheron_1987}. Note that, when constructing local summary statistics (hidden features) in the context of graph convolution, $N \equiv \{(i, j) : j \in \mathcal{N}(i)\}$ where $\mathcal{N}(i)$ denotes the indices of neighbours of $\vec{s}_i$; in this context, the scaling factor $a$ allows for the focusing on information at $\vec{s}_i$ (by increasing $a$) or the information contained in neighbouring nodes (by decreasing $a$). Up to a constant of proportionality, the statistic \eqref{eqn:T} can be made equal to the empirical semivariogram in \eqref{eqn:variogram2} by setting $a=0.5$ and $b=2$ and the empirical madogram (if the data are appropriately transformed beforehand) by setting $a=0.5$ and $b=1$, which is often used when analysing spatial extremes \citep{Cooley2006, Naveau_2009,  Davison_2012_spatial_extremes}. However, it can also represent more general statistics that may be useful when making inference with other non-Gaussian spatial models. We therefore use \eqref{eqn:T} as a basic building block for constructing summary statistics in our GNN architecture.

\section{Neighbourhood definitions}\label{sec:neighbourhoods}

 The definition of the neighbourhood in Equation~(\reff{eqn:propagation2}) of the main text could be important. We consider four possible definitions: the $k$-nearest spatial neighbours for some fixed number $k$; all nodes within a disc of fixed radius $r$; a subset of $k$ neighbours within a disc of fixed radius $r$; and $k$-nearest neighbours subject to a maxmin ordering \citep[e.g.,][]{Guinness_2018}. These definitions are illustrated in Figure~\ref{fig:neighbourhood_definitions}. Several subsampling strategies are possible when choosing a subset of $k$ neighbours within a disc of fixed radius $r$: we use a deterministic algorithm that aims to preserve the distribution of distances within the neighbourhood set, by choosing those nodes with distances to the principal node corresponding to the $\{0, \frac{1}{k}, \frac{2}{k}, \dots, \frac{k-1}{k}, 1\}$ quantiles of the empirical distribution function of distances within the disc. (Note that this subsampling strategy in fact yields up to $k+1$ neighbours for each node, since both the closest and furthest nodes are always included.) 

\begin{figure}[!htb]
 \centering
  \includegraphics[width = \textwidth]{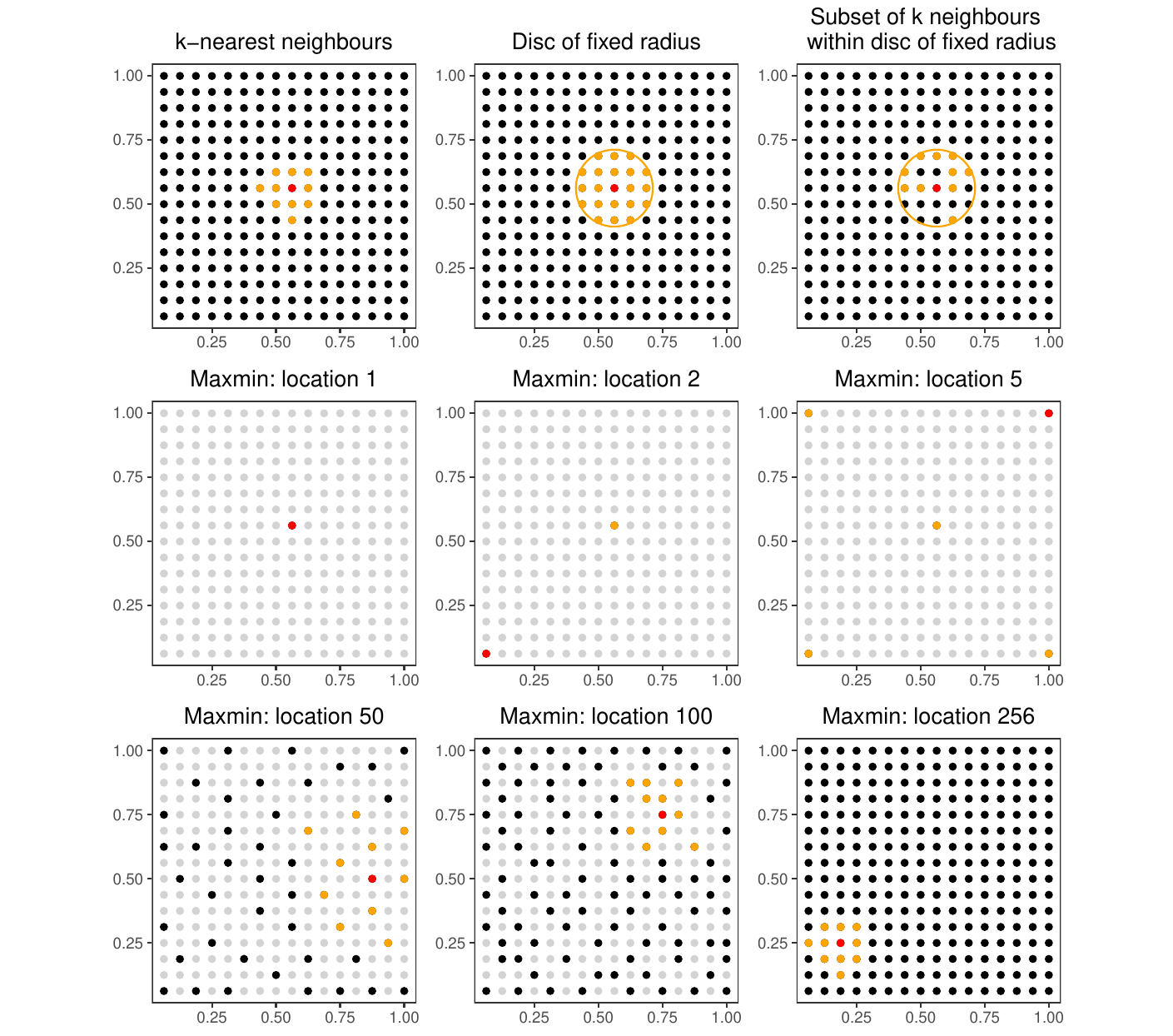}
  \caption{Four definitions for the neighbourhood of a node. (Top row) $k$-nearest neighbours (left), all nodes within a disc of fixed radius $r$ (centre), and a subset of $k$ nodes within a disc of fixed radius $r$ (right), with $k=10$ and $r = 0.1$. In each panel, the principal node, its neighbours, and non-neighbouring nodes are denoted by red, orange, and black points, respectively. (Rows two and three) $k$-nearest neighbours combined with maxmin ordering, where an initial node is selected (location 1), and each subsequent node is selected to maximise the minimum distance to those nodes that have already been selected. In each panel, location $i$ is denoted by a red point; its neighbours, defined as the $k$-nearest nodes from among those that have already appeared in the ordering, are denoted by orange points; nodes that are not neighbours but precede the $i$th point in the ordering are denoted by black points; and nodes that appear after the $i$th point in the ordering are denoted by grey points.  
  }\label{fig:neighbourhood_definitions} 
\end{figure}

Before proceeding to an empirical analysis, we first discuss several intrinsic properties of these neighbourhood definitions. First, with a fixed bounded spatial domain, the disc-of-fixed-radius definition results in a computational complexity of $\mathcal{O}(r^2n^2)$, since increasing the number $n$ of data points simultaneously increases the total number of convolutions that must be performed, and the number of neighbours for each node. The other definitions that we consider have a computational complexity of $\mathcal{O}(kn)$. Note that this difference in computational complexity does not necessarily translate into a meaningful difference in runtime, since the computations involved with GNNs are done in parallel on GPUs containing thousands of cores. Second, choosing a subset of $k$ neighbours within each disc requires the specification of two hyperparameters ($k$ and $r$). Third, under disc-based definitions, it is possible for a node to be disconnected if no other nodes fall within its neighbourhood disc (the likelihood of this occurring decreases with increasing disc radius $r$). Fourth, with a fixed bounded spatial domain, the distance between $k$ nearest neighbours tends to zero as $n$ becomes large, and this could compromise the estimators ability to properly model medium-to-long-range spatial dependencies; the use of a maxmin ordering can overcome this limitation, since it promotes connectivity between points that are not necessarily in close proximity to each other. Finally, to avoid extrapolation when applying estimators to larger data sets than those used during training (see, e.g., Section~\reffmain{sec:application}), it may be necessary to define neighbourhoods that maintain the distribution of distances between nodes and their neighbours as the sample size $n$ increases (e.g., disc-of-fixed-radius definitions). Although it is helpful to bear these properties in mind when constructing an estimator, they do not always translate into meaningful differences, as we illustrate in the following sensitivity analysis. 

We now conduct an experiment to investigate empirically the effect of the neighbourhood definitions described above. We construct a range of GNN-based estimators, each differing only by the neighbourhood definition and specific choice of hyperparameters. We consider the Gaussian process model described in Section~\reff{sec:GP}, with spatial configurations sampled from the Matérn cluster process described in the main text. Figure~\ref{fig:neighbourhoods}, columns one and two, shows the empirical RMSE and the post-training inference time against the respective hyperparameters.  The estimators perform similarly well with respect to RMSE except for disc-of-fixed-radius definitions with very small hyperparameter choices. The estimators also have similar run-times since, although the number of computations increases linearly with the number of neighbours, the computations are done in parallel, as discussed above. Figure~\ref{fig:neighbourhoods}, column three, shows the empirical RMSE and the post-training inference time against the sample size $n$, for each neighbourhood definition and with the hyperparameter(s) selected to those values with minimum RMSE in columns one and two. Again, the estimators perform similarly well in terms of RMSE, and are able to extrapolate to larger sample sizes than those used during training. 

\begin{figure*}[t!]
    \centering
    \includegraphics[width = \textwidth]{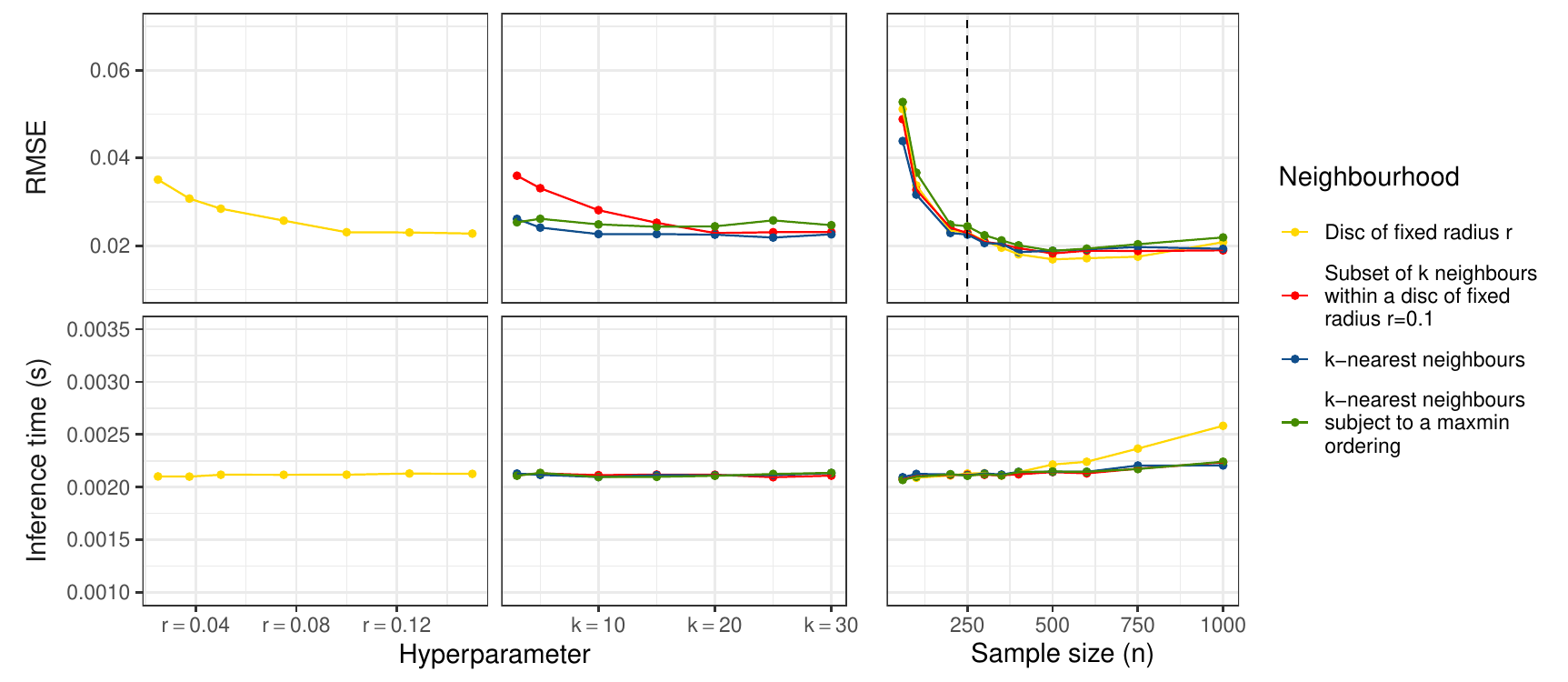}  
    \caption{
     The empirical RMSE (first row) and the single-data-set inference time (second row), against the hyperparameter ($r$ or $k$; first and second columns) or sample size $n$ (third column) for several GNN-based estimators. The estimators differ in the way the neighbours of a node are defined: all nodes within a disc of fixed radius $r$, a subset of $k$ neighbours within a disc of fixed radius $r=0.1$, $k$-nearest neighbours, and $k$-nearest neighbours subject to a maxmin ordering (see Figure~\ref{fig:neighbourhood_definitions}). In the top-right panel, a dashed line is used to denote the sample size used during training (results to the right of this line correspond to extrapolation to larger sample sizes).  
    }\label{fig:neighbourhoods}
\end{figure*} 

 Overall, in this experiment, the proposed estimator appears to be relatively insensitive to the choice of neighbourhood definition and hyperparameters. Although the estimator is relatively insensitive in this experiment, the results could vary depending on the context and model being fitted, and in certain situations it may be necessary to tune the neighbourhood hyperparameter(s) to achieve optimal results.

\section{Probability distribution for the spatial locations $S$}\label{sec:experiment:variablesamplesizes}

Our proposed methodology differs to many other approaches in that, to facilitate amortised inference whereby the estimator is constructed before data have been collected, it is often necessary to define a distribution $\Omega(\cdot)$ for the spatial locations $S$. In this section, we investigate several properties of our methodology with respect to this distribution. 

\paragraph{Variable numbers of spatial locations} As discussed in the main text, a GNN-based neural Bayes estimator can be applied to data collected over any set of spatial locations, $S$, and with any number of locations, $n$. However, Bayes estimators are generally a function of $n$, and this must be accounted for during training if the estimator is to generalise over a wide range of possible sample sizes. To illustrate this property, we train three GNN-based estimators for the Gaussian process model of Section~\reff{sec:GP} with different distributions for $S$. We train the first and second estimators with data sets containing exactly $30$ or $1000$ sampled locations, respectively, and we train the third estimator with $n$ treated as a discrete uniform random variable with support between 30 and 1000 inclusive, so that it is trained with a range of sample sizes. Irrespective of $n$, the spatial locations are sampled from a uniform binomial point process \citep[pg.~59]{Illian_2008_point_patterns}, which simply consists of $n$ points randomly scattered in the unit square; we denote this point process by $\text{UBPP}(n)$. Note that here we adopt a uniform binomial point process so that we can specify the exact number $n$ of spatial locations in each realisation (many point processes, e.g., the Matérn cluster process, only allow one to specify the expected number of spatial locations in each realisation).

 Figure~\ref{fig:priorS}, left panel, shows the empirical RMSE for each estimator against the number of spatial locations, $n$. The estimators trained with fixed $n$ perform reasonably well when $n$ is close to the corresponding value used during training, but poorly for other sample sizes. On the other hand, the estimator trained with a range of sample sizes performs well in all cases: this behaviour is expected from Theorem~\ref{thm:GNNproof} in Appendix~\reffmain{app:Proof:priormeasureS}. 
 
\begin{figure*}[t!]
    \centering
    \includegraphics[width = \textwidth]{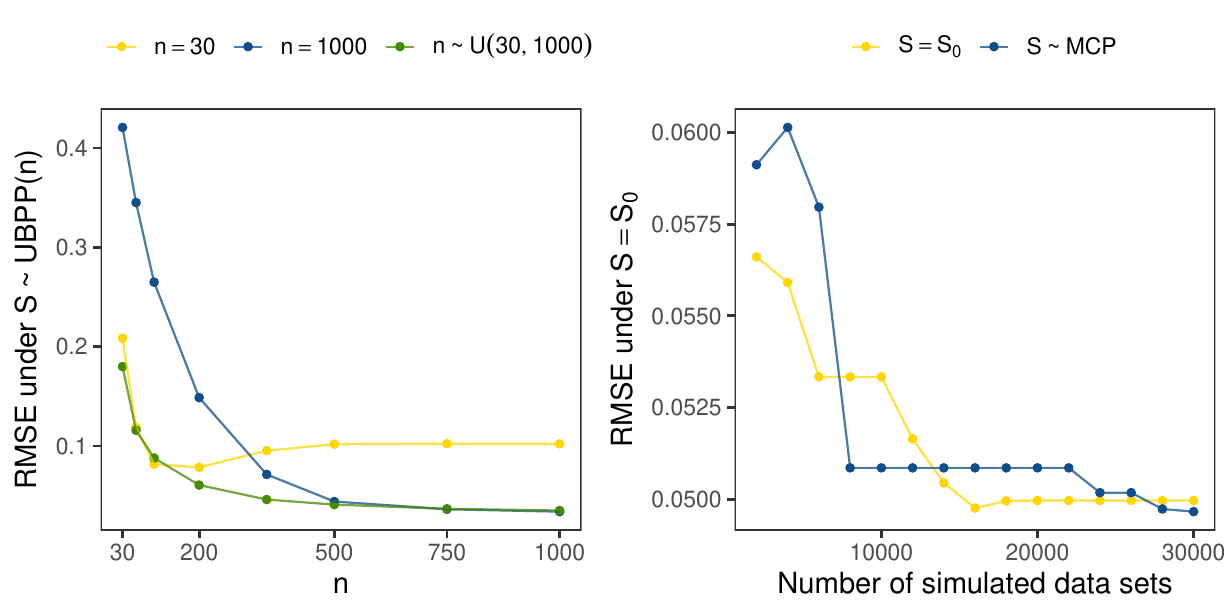}  
    \caption{(Left) The empirical RMSE against the number of spatial locations, $n$, for three GNN-based estimators trained with $S\sim \text{UBPP}(n)$, where $n$ is fixed to $30$, fixed to $1000$, or sampled uniformly between $30$ and $1000$. (Right) The empirical RMSE against the number of simulated data sets used to train two GNN-based estimators: the first with $S = S_0$ fixed, where $S_0 \sim \text{UBPP}(250)$; and the second with $S$ random and following a Matérn cluster process (MCP).  
    }\label{fig:priorS}
\end{figure*}

\paragraph{Simulation efficiency with random $S$} A possible concern when treating $S$ as random during the training stage is that one may require many more simulations to achieve a similar level of accuracy with respect to a specific set of locations, $S_0$, compared with an estimator trained with $S = S_0$ fixed. However, we do not find this to be the case in our experiments. 
 
Consider the following experiment. First, we train an estimator with $S = S_0$ fixed, where $S_0 \sim \text{UBPP}(250)$. Then, we train a second estimator with $S$ random and following the Matérn cluster process described in Section~\reffmain{sec:SimulationIntro}. Realisations from this cluster process vary from highly clustered to approximately uniform (recall Figure~\reffmain{fig:spatialpatterns}), and one might therefore expect that, when assessed with respect to $S_0$, many more simulations would be required to achieve a similar performance to the first estimator. However, Figure~\ref{fig:priorS}, right panel, shows that the empirical RMSE, computed with respect to $S_0$ (i.e., using simulated data in which all sets of spatial locations are fixed to $S_0$), decreases at a similar rate for both estimators. 
 
\section{Additional figures and tables}\label{sec:additionalfigures}

\begin{figure*}[!htb]
    \centering
    \includegraphics[width = \textwidth]{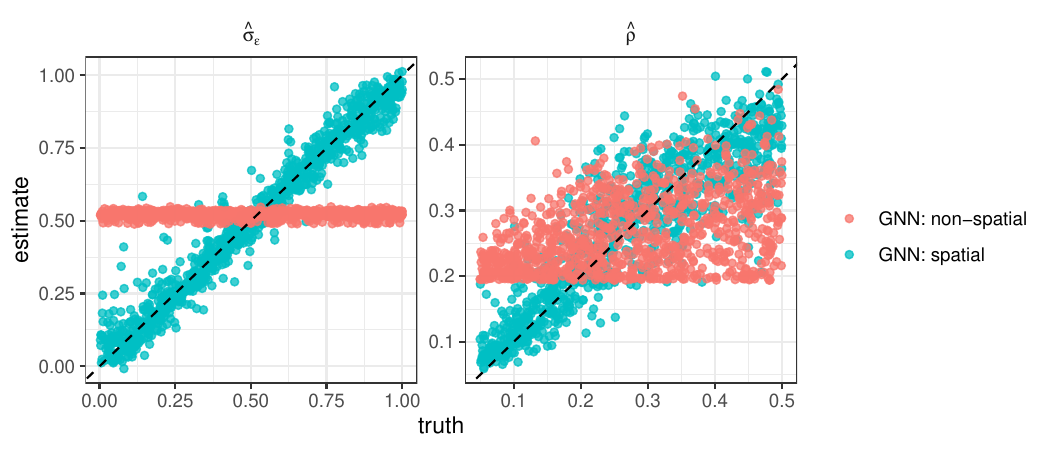}  
    \caption{Parameter estimates against true values from 1000 data sets, each with the same $n = 256$ spatial locations representing a regular $16\times 16$ grid over the unit square, from two GNN-based estimators for the Gaussian process model of Section~\ref{sec:GP}. The estimators differ only in their definition of the propagation modules; a ``spatial'' version given by Equations (\reff{eqn:propagation1})--(\reff{eqn:propagation3}) of the main text, and a ``non-spatial'' version that omits the spatial weighting function $\vec{w}(\cdot, \cdot)$. The spatial GNN estimator clearly outperforms its non-spatial counterpart. 
    }\label{fig:spatial_vs_nonspatial}
\end{figure*}

\begin{figure}[t!]
\begin{center}
  \begin{tikzpicture}[scale=1,every node/.style={transform shape}]
[line width=1pt]
\path (-2.5,3) node (Y1) {$\{\boldsymbol{Z}_1, S_1\}$};
\path (-2.5,1.5) node (dots1) [shape=circle,draw = none, minimum size=1cm] {$\vdots$};
\path (-2.5,0) node (Yn) {$\{\boldsymbol{Z}_m, S_m\}$};

\path (0,3) node (g1) [shape=rectangle,minimum size=1cm,draw] {$\vec{\psi}(\cdot)$};
\path (0,1.5) node (dots2) [shape=rectangle, draw=none,minimum size=1cm] {$\vdots$};
\path (0,0) node (gn) [shape=rectangle,minimum size=1cm,draw] {$\vec{\psi}(\cdot)$};

\path (2,3) node (V1) [shape=circle,minimum size=1cm,draw] {$\vec{R}_{1}$};
\path (2,1.5) node (dots3) [shape=circle, draw=none,minimum size=1cm] {$\vdots$};
\path (2,0) node (Vn) [shape=circle,minimum size=1cm,draw] {$\vec{R}_{m}$};

\path (4,1.5) node (a) [shape=rectangle,minimum size=1cm,draw] {$\vec{a}(\cdot)$};

\path (5.75,1.5) node (T) [shape=circle,minimum size=1cm,draw] {$\vec{T}$};

\path (7.5,1.5) node (h) [shape=rectangle,minimum size=1cm,draw] {$\vec{\phi}(\cdot)$};

\path (9.25,1.5) node (thetahat) [shape=circle,minimum size=1cm,draw] {$\hat{\vec{\theta}}$};

\draw [->] (Y1) to (g1);
\draw [->] (Yn) to (gn);
\draw [->] (g1) to (V1);
\draw [->] (gn) to (Vn);
\draw [->] (V1) to (a);
\draw [->] (Vn) to (a);
\draw [->] (a) to (T);
\draw [->] (T) to (h);
\draw [->] (h) to (thetahat);

\end{tikzpicture}
\caption{The structure of a GNN-based neural Bayes estimator for making inference from $m$ mutually independent replicates, $\vec{Z}_1,\dots, \vec{Z}_m$, with associated spatial locations, $S_1, \dots, S_m$. The replicates are first processed independently by the propagation and readout modules described in Section~\reffmain{sec:GNN:GeneralFramework} (this operation is denoted by $\vec{\psi}(\cdot)$ in this schematic), which yields fixed-length summary statistics, $\vec{R}_1,\dots, \vec{R}_m$. These summary statistics are aggregated using a permutation-invariant set function, $\vec{a}(\cdot)$, into a single vector of summary statistics, $\vec{T}$, which is then transformed by an MLP $\vec{\phi}(\cdot)$ into parameter estimates $\hat{\vec{\theta}}$.  
}\label{fig:GNN-estimator:DeepSets}
\end{center}
\end{figure}
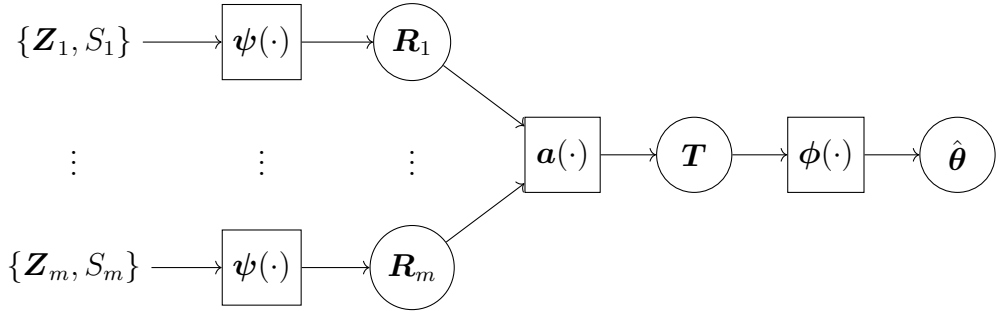

\begin{figure}[!htb]
 \centering
  \includegraphics[width = \textwidth]{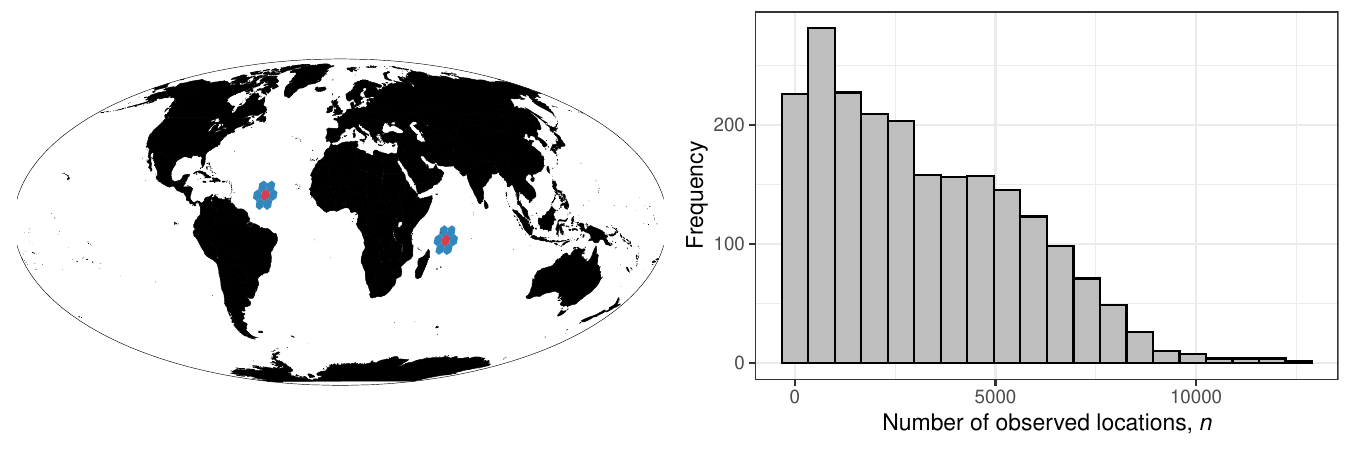}
  \caption{(Left) Two cell clusters used in the application study of Section~\reff{sec:application} of the main text; the parameter estimates for a given cell (red) are obtained using both the data within that cell and the data within its neighbouring cells (blue). (Right) Histogram of the number of observations, $n$, for all cell clusters used in the application study of Section~\reff{sec:application} of the main text.  
   }\label{fig:SST:clustering}
\end{figure}
 
\begin{figure*}[t!]
    \centering
    \includegraphics[width = \textwidth]{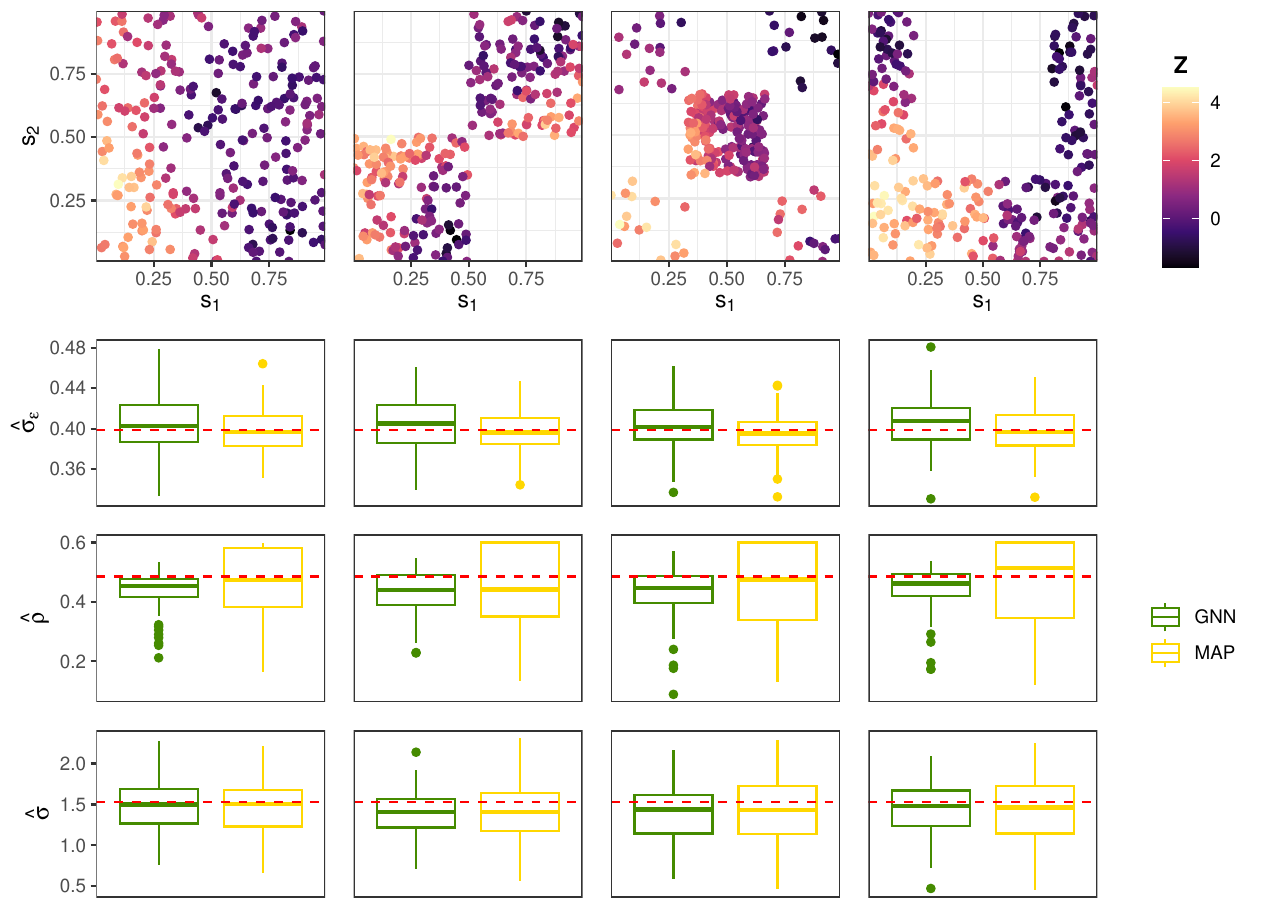}  
    \caption{Several spatial data sets (top row), each with $n = 250$ spatial locations, and empirical marginal sampling distributions (second, third, and fourth rows) of two estimators for the Gaussian process model of Section~\ref{sec:application} with parameters denoted by the dashed line. The estimators are the MAP estimator and a GNN-based neural Bayes estimator. A single GNN was trained for all data sets. Our neural credible-intervals for $\rho$, $\sigma$, and $\sigma_\epsilon$ were found to have empirical coverages of 95.2\%, 94.1\%, and 95.1\%, respectively, which are close to the nominal value of 95\%.    
    }\label{fig:GPSigmaVaried}
\end{figure*}

\begin{figure}[!htb]
	\vspace{1cm}
	    \centering
  \includegraphics[width = 0.9\textwidth]{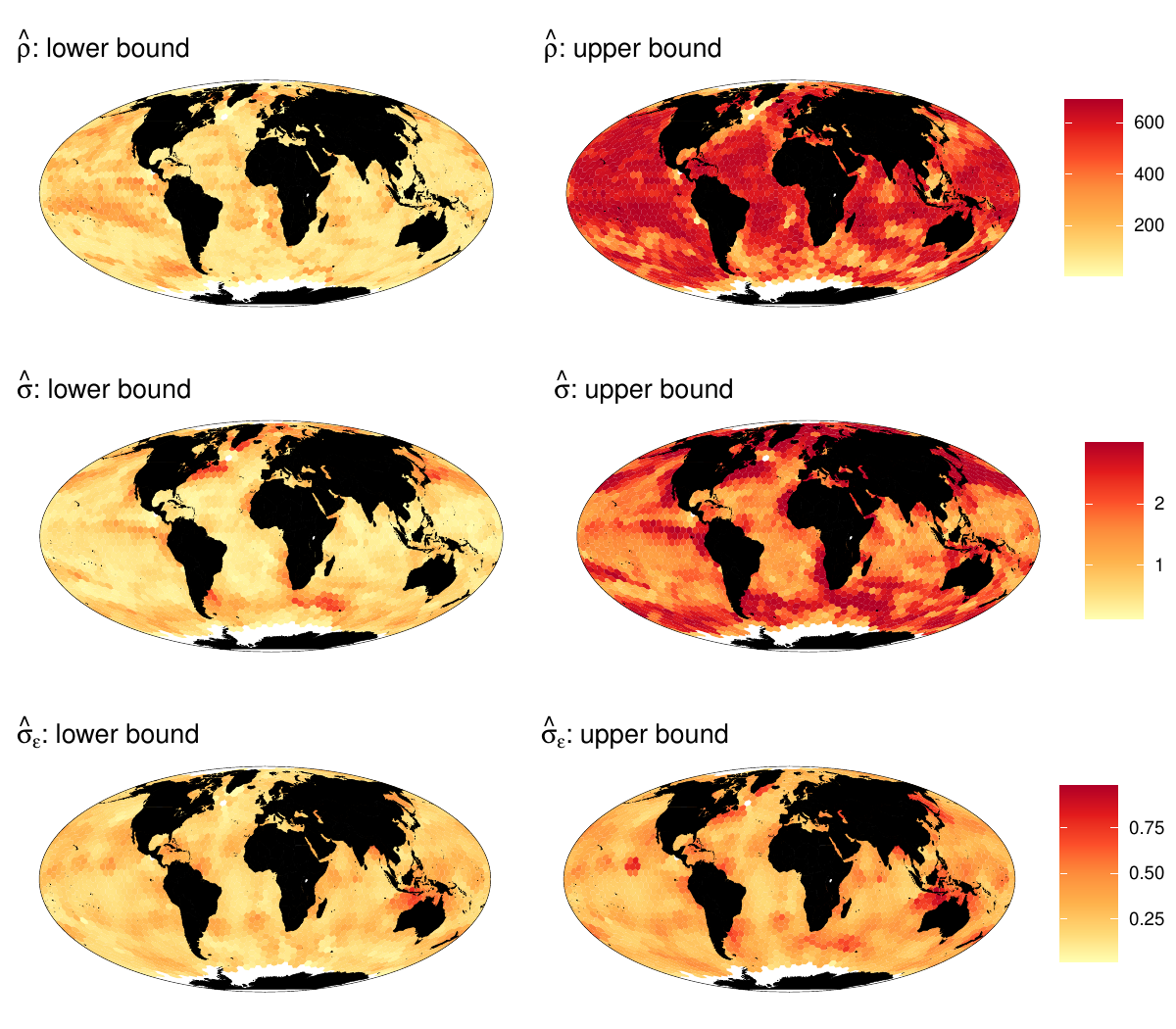}
  \caption{  
Spatially varying estimates of the marginal 0.025 quantile (left column) and marginal 0.975 quantile (right column) denoted as lower and upper bounds, respectively, for each parameter of the Gaussian process model used in Section~\reff{sec:application} of the main text. The first, second, and third rows correspond to the range parameter, $\rho$, process standard deviation, $\sigma$, and measurement-error standard deviation, $\sigma_\epsilon$, respectively. The globe is partitioned using the ISEA Aperture 3 Hexagon (ISEA3H) discrete global grid (DGG) at resolution 5. 
  }\label{fig:SST:intervals}
\end{figure}

\begin{figure}[!htb]
	\vspace{1cm}
	    \centering
  \includegraphics[width = \textwidth]{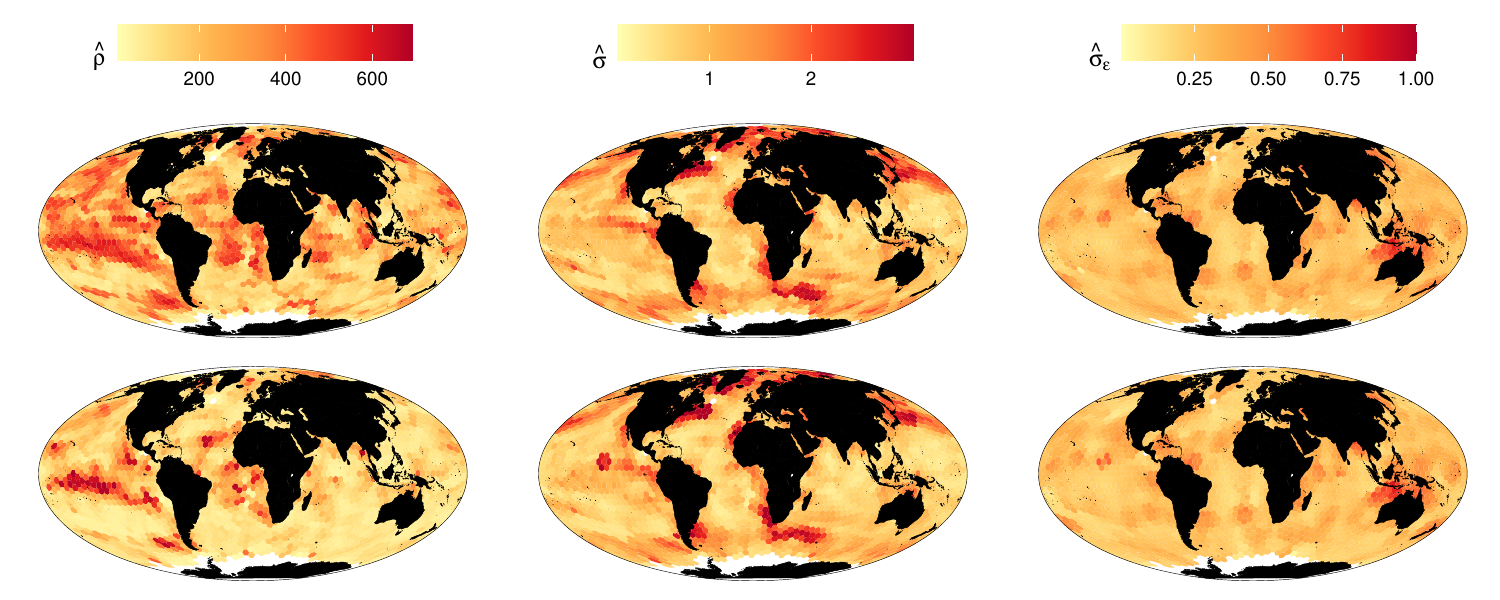}
  \caption{  
Spatially varying point estimates obtained using a GNN-based neural Bayes estimator (top row) and the MAP estimator (bottom row) for each parameter of the Gaussian process model used in Section~\reff{sec:application} of the main text. The first, second, and third columns correspond to the range parameter, $\rho$, process standard deviation, $\sigma$, and measurement-error standard deviation, $\sigma_\epsilon$, respectively. The globe is partitioned using the ISEA Aperture 3 Hexagon (ISEA3H) discrete global grid (DGG) at resolution 5. Recall that for computational reasons, the MAP estimates are capped to 3000 data points per region.
  }\label{fig:SST:GNN_ML_estimates}
\end{figure}

\clearpage
\bibliographystyle{apalike} 
\putbib[bibliography]
\end{bibunit}

}{}

\end{document}